\documentclass{article}
\usepackage{amsmath,amssymb,graphicx}
\usepackage{hyperref}

\def\be{\begin{eqnarray}}
\def\ee{\end{eqnarray}}

\def\l[{\phantom.[}

\newcommand{\ch}{\mathop{\mathrm{ch}}\nolimits}

 \newcommand{\Res}{\mathop{\mathrm{Res}}\limits}

\hoffset= - 1.0in         

\begin{document}

\title{\vspace{1cm}{\Large {\bf Spectral duality in elliptic systems, six-dimensional gauge
    theories and topological strings}\vspace{.2cm}}
\author{{\bf A. Mironov$^{a,b,c,d,}$}\footnote{mironov@lpi.ru; mironov@itep.ru}, \ {\bf A. Morozov$^{b,c,d,}$}\thanks{morozov@itep.ru}, \ and \ {\bf Y. Zenkevich$^{b,d,e,}$}\thanks{yegor.zenkevich@gmail.com}}
\date{ }
}

\maketitle

\vspace{-6cm}

\begin{center}
\hfill FIAN/TD-03/16\\
\hfill IITP/TH-03/16\\
\hfill ITEP/TH-04/16\\
\hfill INR-TH-2016-007
\end{center}

\vspace{4.2cm}

\begin{center}
$^a$ {\small {\it Lebedev Physics Institute, Moscow 119991, Russia}}\\
$^b$ {\small {\it ITEP, Moscow 117218, Russia}}\\
$^c$ {\small {\it Institute for Information Transmission Problems, Moscow 127994, Russia}}\\
$^d$ {\small {\it National Research Nuclear University MEPhI, Moscow 115409, Russia }}\\
$^e$ {\small {\it Institute of Nuclear Research, Moscow 117312, Russia }}
\end{center}

\vspace{1cm}

\begin{abstract}
We consider Dotsenko-Fateev matrix models associated with compactified Calabi-Yau threefolds. They can be constructed with the help of explicit expressions for refined topological vertex, i.e. are directly related to the corresponding topological string amplitudes. We describe a correspondence between these amplitudes, elliptic and affine type Selberg integrals and gauge theories in five and six dimensions with various matter content. We show that the theories of this type are connected by spectral dualities, which can be also seen at the level of elliptic Seiberg-Witten integrable systems. The most interesting are the spectral duality between the XYZ spin chain and the Ruijsenaars system, which is further lifted to self-duality of the double elliptic system.
\end{abstract}

\vspace{.5cm}


\tableofcontents

\vspace{2cm}

\section{Introduction}

Gauge theories with eight supercharges in 4d, 5d and 6d can be
effectively analyzed from the string theory perspective. This view is
natural if one wants to study deformations of these theories and
understand their structure in geometric terms. Also this family of
gauge theories turns out to be a focus point of several dualities, some
of which are still in need of a full explanation.

One of these dualities is the AGT correspondence, relating partition
functions of gauge theories to 2d CFT conformal blocks. It was first
observed for 4d theories~\cite{AGT}, then generalized to
5d~\cite{5dAGT} and very recently to 6d~\cite{6dAGT}. The AGT relation
was motivated by the study of the worldvolume theory on the stack of
M5 branes wrapping a Riemann surface, the bare spectral
curve~\cite{Gaiotto}. This geometric point of view naturally
incorporated various properties of 4d $\mathcal{N}=2$ gauge theories:
$S$-duality, Seiberg-Witten curve, BPS states counting and relation
with integrable systems. Any geometric meaning of the AGT duality for
5d gauge theories, which features $q$-deformed 2d CFT, is not so
manifest.

Gauge theories in five dimensions can be obtained using a different
approach, the geometric engineering technique, which relates them
to type IIB strings on the $(p,q)$-brane web~\cite{Aharony:1997bh,GGM2} or
topological strings on toric Calabi-Yau three-folds. This approach
allows for direct computation using the refined topological vertex
technique \cite{ref,IKV}, and explicitly reproduces the Nekrasov partition function of
gauge theories.

$(p,q)$-webs and toric CY backgrounds have a natural symmetry, the
spectral~\cite{spedu,MMZZ}, or fiber-base, duality, which is also the
$S$-duality of IIB strings. This duality connects gauge theories with
different gauge groups and matter content, which, however, have the
same partition functions and the same set of BPS particles: the
instantons and $W$-bosons get exchanged. The spectral duality has been
studied for linear quiver gauge theories~\cite{Bao} and for $SU(2)$
gauge theories with $N_f \leq 8$ fundamental matter
multiplets~\cite{enhanced}.

It has also been understood that the spectral duality is closely
related to the AGT duality. In~\cite{Aganagic:2013tta} (see also \cite{Sulk}) the Dotsenko-Fateev (DF)
integrals for conformal blocks of the $q$-deformed CFT~\cite{FDagt,DV,MMSS5} have been
rewritten as sums over residues, each corresponding to a fixed point
in the instanton moduli space of a 5d gauge theory. However, this
gauge theory turned out to be not the theory related to the conformal
block by the AGT duality, but rather its spectral dual. Thus, the
AGT relation is obtained as action of the spectral duality on the
DF integrals of the $q$-deformed CFT.

In~\cite{genMD},~\cite{Morozov:2015xya},~\cite{MMZ} we initiated a
program to better understand the relationship between 5d gauge
theories, $q$-deformed conformal blocks and structure of refined
topological string amplitudes on toric CY. Here we would like to
extend our analysis to the compactified toric CY, which correspond to 6d
gauge theories and to the elliptic deformation of the conformal algebra.

In the spirit of~\cite{Morozov:2015xya}, we show how to combine the
elementary building blocks, amplitudes of the refined topological
string in order to obtain the measures of the DF-type integral
representations of conformal blocks in various theories. In this way,
we obtain the integrals for $W_N$ algebras (also known as $A_N$
$q$-Selberg integrals), toric $q$-deformed conformal blocks (featuring
in the ``elliptic AGT relations''~\cite{6dAGT}), conformal blocks
corresponding to affine root systems (related to the 6d gauge
theories), and the setup corresponding to the 6d theory with
adjoint matter.

It is well-known that gauge theories we are considering correspond to
the Seiberg-Witten integrable systems \cite{GKMMM}-\cite{GM},\cite{GGM2}. In the case of compactified brane
diagram, these integrable systems are of elliptic type. They include
the elliptic Ruijsenaars, the XYZ spin chain and the still mysterious double
elliptic integrable system~\cite{Dell}. We propose the spectral
duality for these systems and make a few qualitative tests of it.

One motivation for going to 6d is to probe the $(2,0)$ 6d
superconformal theory, which is thought to originate from coincident
M5 branes. However, what we actually obtain is the $T$-dual theory,
the $(1,1)$ 6d gauge theory. This is the dimensional reduction of the
10d minimal supersymmetric gauge theory, which upon further reduction
gives the $\mathcal{N}=2$ theory in 5d and $\mathcal{N}=4$ theory in
4d. The $(1,1)$ gauge theory can also be thought of as a K\"ahler
gravity theory, which is the target space description of the
microscopic ``quantum foam'' geometry of topological
strings~\cite{Iqbal:2003ds}. The theory can be put in the 6d
$\Omega$-background with \emph{three} equivariant parameters
$q_{1,2,3}$ corresponding to three elements of the Cartan subalgebra
of $SO(6)$. The ``instantons'' of the 6d theory, fixed under these
isometries are identified with the atoms of the melting crystal model,
and the partition function of the gauge theory is identified with the
index version of the topological vertex.

We start our discussion by reminding the brane descriptions of gauge
theories in 4d, 5d, 6d and their relation to (refined) topological
strings. We then provide an exhaustive list of possible topological
string amplitudes (including compactified toric diagrams), which are
suitable as building blocks for quiver gauge theories in 5d and 6d. We
also generalize the dictionary obtained in~\cite{Morozov:2015xya} to
these amplitudes and describe the corresponding Dotsenko-Fateev type
integrals. Our approach to the DF integrals exploits also the ``triality''
between 5d gauge theories, 3d theories and $q$-deformed conformal
blocks proposed in~\cite{Aganagic:2013tta}. We interpret the sums
over Young diagrams as the sums over residues and then investigate
structure of the corresponding integral.

In the second part, we focus on one particular amplitude for the
compactified CY, which corresponds to the 5d gauge theory with adjoint
matter, or, employing the spectral duality, to the 6d linear quiver of
$U(1)$ groups. We find here a close counterpart of the triality, which gives
the \emph{affine} version of the $q$-Selberg integrals. Investigating
the pole structure of these integrals we encounter a new interesting
phenomenon: the contour of integration does not encircle all the poles
of the integrand. The meaning of the missing poles turns out to be
quite remarkable: they correspond to instantons of the
\emph{six-dimensional} gauge theory enumerated by the plane partitions
(3d Young diagrams). Even more remarkable is the fact that each
residue exactly reproduces the equivariant $K$-theory index of the
corresponding fixed point in the instanton moduli space of the
$\mathcal{N}=(1,1)$ 6d theory~\cite{Nekr-QUARKS, Nekrasov:2014nea}. We
also point out that the \emph{affine} Selberg integrals and their
generalizations, which we introduce, have a very nice cohomology
limit, in which they turn into the 6d counterparts of the LMNS
integrals \cite{LMNS}. Taking a further limit reduces the integrals to the
standard 4d LMNS ones. Thus, our extended integral provides a simple
explanation for the AGT relation between the two very different
integral representations of the same quantity: the DF integrals
(representing the conformal blocks) and the LMNS integrals (computing
the Nekrasov partition function).

In the last part of the paper, we discuss the Seiberg-Witten
integrable systems, corresponding to the gauge theories we have
considered. We first recall the general idea of the integrable system
construction, and then proceed to study action of the spectral duality
on integrable systems \emph{a la}~\cite{spedu}. We generalize the
results known for the rational and trigonometric systems to the elliptic ones:
the elliptic Ruijsenaars model, the XYZ spin chain, the double elliptic systems
and their generalizations. We show that the spectral duality originating
from the $(p,q)$-brane rotation indeed gives a nontrivial identification
between different elliptic systems.

\subsection{Brane pictures\label{1.1}}

\begin{figure}[h]
  \centering
\includegraphics[width=12cm]{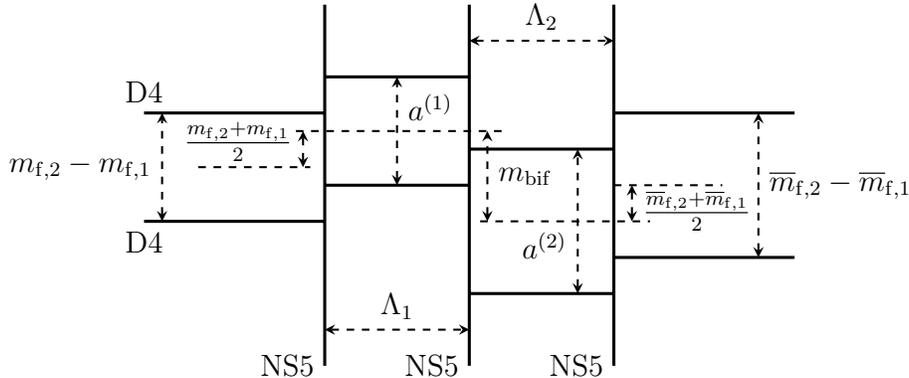}
  \caption{Type IIA brane diagram corresponding to $U(2)^2$ linear
    quiver.}
  \label{fig:1}
\end{figure}

For 4d theories the relevant brane construction is provided by Type
IIA theory~\cite{Witten:1997sc} (see also related subjects in \cite{Witmore}). To obtain a linear $U(N)^{M-1}$ gauge
theory one considers a set of $M$ ``vertical'' NS5 branes extending in
the $(x^0,x^1,x^2,x^3,x^4,x^5)$ directions and $N$ ``horizontal'' D4
branes suspended between them in the directions
$(x^0,x^1,x^2,x^3,x^6)$. The $(x^5,x^6)$ projection of this setting is
shown in Fig.~\ref{fig:1}. On the segments of D4 branes suspended
between each pair of adjacent NS5 branes lives a gauge theory with
$U(N)$ gauge group, which is spontaneously broken down to $U(1)^N$ by
the adjoint scalar vacuum averages $a^{(\alpha)}_i$. These averages
are represented by the vertical distances between the D4 branes. The
(asymptotic) distance between the two NS5 branes represent the
complexified coupling constant $\Lambda_{\alpha} = -i \tau_{\alpha} =
\frac{4\pi}{g^2_{\alpha}} - \frac{i \theta_{\alpha}}{2\pi}$ of the
gauge group. The neighbouring gauge groups in the linear quiver are
coupled through a bifundamental hypermultiplet. Its mass
$m_{\mathrm{bif}}$ is given by the relative positions of the centers
of masses of D4 branes to the left and to the right of the
corresponding NS5 brane. There are $N$ semi-infinite D4 branes coming
from the left --- they correspond to $N$ fundamental matter fields
with masses $m_{\mathrm{f},i}$ coupled to the first $U(N)$ gauge
factor of the quiver. These branes can also be understood as arising
from an additional gauge theory with vanishing coupling
constant. Similarly $N$ branes extending to the right of the diagram
correspond to $N$ antifundamental hypermultiplets with masses
$\overline{m}_{\mathrm{f},i}$ coupled to the last gauge factor. One
can make the hypermultiplets infinitely massive while simultaneously
sending the gauge coupling to zero. This corresponds to the picture,
where all the semi-infinite D4 branes are moved infinitely high or
low. The D4 branes have tension, which bends the NS5 branes, so they
are no longer asymptotically parallel to each other. This simply means
that there is no well-defined coupling at high energies, and the
theory is asymptotically free. Indeed, bending of NS5 branes can be
found by solving for a minimal surface in 3d space spanned by $(x^4,
x^5, x^6)$. Thus, the brane diagram in Fig.~\ref{fig:1} is only
schematic --- NS5 branes should be viewed as curved surfaces pulled by
D4 branes. Away from D4 branes one gets $x^6(x^4, x^5) \sim \ln |x^4 +
i x^5|$ --- characteristic behavior of an asymptotically free
theory. If there are the same numbers of D4 branes pulling an NS5
brane to the left and to the right, then the tension is balanced and
asymptotically one has $x^6(x^4, x^5) \sim \mathrm{const}$, so that
the resulting gauge theory is conformal in the UV.

To analyze the non-perturbative effects, such as instantons, in the
gauge theory the picture has to be lifted to M-theory. The extra
coordinate forms a circle $S^1_{R_{11}}$, which radius is proportional
to Type IIA string coupling constant. Then, the NS5 and D4 branes are
described by a single M5-brane, wrapping the complex Seiberg-Witten
curve $\Sigma$, which should be holomorphic due to supersymmetry. The
matter content, prepotential, BPS spectrum and global symmetries can
be read off from the curve and the corresponding brane diagram. In
particular, M2 branes suspended between the sheets of $\Sigma$ and
wrapping special contours on it correspond to BPS particles in the
gauge theory. Partition function of the gauge theory can be found by
evaluating the periods of the Seiberg-Witten differential $\lambda$
over the Seiberg-Witten curve $\Sigma$.

Let us describe one more system associated with the brane picture: the
Seiberg-Witten integrable system. The phase space of this system can
be understood as follows. The family of Seiberg-Witten curves is
parameterized by the Coulomb moduli of the gauge theory, which
correspond to positions of the D4 branes in the Type IIA
picture. These are the integrals of motion, or Hamiltonians of the
integrable system. Thus, the Seiberg-Witten curve is identified with
the spectral curve of the integrable system, which is the generating
function of the Hamiltonians. For fixed values of the Hamiltonians,
the system moves on a torus. This torus is given by the Jacobian of
the Seiberg-Witten curve, so that the whole phase space is a torus
fibration over the Coulomb branch of the gauge theory moduli
space. The surprising fact is that Seiberg-Witten integrable systems
can in fact be described very concretely and coincide with some
classic integrable systems: e.g.\ for 4d $U(N)^{M-1}$ linear quiver it
is given by the periodic $\mathfrak{gl}_M$ XXX spin chain with $N$
spins \cite{GGM1}. The masses of the hypermultiplets correspond to Casimir
operators of the spins and the couplings enter the twist matrix of the
chain.

\begin{figure}[h]
  \centering
\includegraphics[width=5cm]{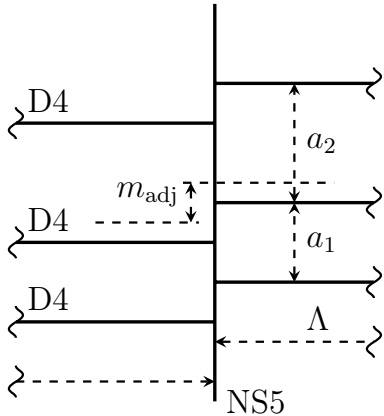}
\caption{Type IIA brane diagram corresponding to $U(3)$ theory with
  adjoint multiplet. Notice the cyclic identification of the lines on
  the left and right.}
  \label{fig:10}
\end{figure}

In a similar way one can consider gauge theories with adjoint
matter. In this case the $x^6$ direction (horizontal in
Fig.~\ref{fig:1}) should be compactified on a circle $S^1_{R_6}$, so
that the D4 branes no longer extend to infinity on the left and right
margins of the page, but wrap the $S^1_{R_6}$ circle with both ends
attached to a single NS5 brane. The resulting arrangement is shown in
Fig.~\ref{fig:10}. By compactifying diagrams with more NS5 branes, one
gets a ``necklace'' quiver of gauge groups coupled by
bifundamentals. For the resulting gauge theory to be well defined in
the UV one has to make all the gaps between NS5 branes expanding
asymptotically, or staying constant. The only way to actually obey
this constraint is to make the number of D4 branes suspended between
each two NS5 branes the same, which gives the $U(N)^M$ quiver. The
integrable system corresponding to the 4d $U(N)$ theory with adjoint
matter is $N$-particle elliptic Calogero model \cite{SWCal}, and for necklace
quivers it is a certain multipoint generalization of Calogero
system \cite{MMZZ}. The mass of the adjoint multiplet corresponds to the Calogero
coupling constant and the \emph{gauge theory} coupling constant is
encoded in the elliptic parameter of the system.

\begin{figure}[h]
  \centering
\includegraphics[width=10cm]{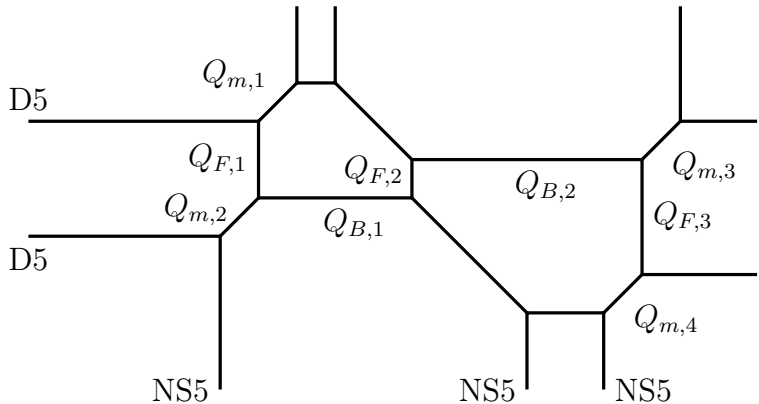}
\caption{Type IIB $(p,q)$-brane web, or CY toric diagram,
  corresponding to 5d $U(2)^2$ linear quiver.}
  \label{fig:2}
\end{figure}

One can also compactify the $x^4$ direction (which is perpendicular to
the plane of Figs.~\ref{fig:1}, \ref{fig:10}). To get the gauge theory
interpretation of this setup we perform $T$-duality along the
resulting circle $S^1_{1/R_4}$. Under $T$-duality Type IIA theory
turns into Type IIB, which has odd instead of even D-branes. Thus,
after $T$-duality, D4 branes, which were suspended between the NS5
branes and spanned $(x^0,x^1,x^2,x^3,x^6)$, being transverse to
$S^1_{1/R_4}$, become D5-branes, wrapped over $S^1_{R_4}$ and spanning
$(x^0,x^1,x^2,x^3,x^4,x^6)$. The worldvolume theory on D5 branes is
generally a six-dimensional gauge theory. However, in the $x^6$
direction the brane has a finite span $\Lambda$, so the resulting low
energy theory is five-dimensional gauge theory on
$(x^0,x^1,x^2,x^3,x^4) \in \mathbb{R}^4 \times S^1_{R_4}$. However, as
in the NS5-D4 system discussed above, we have to include brane tension
in our picture. The difference with the D4 case is that now we have to
solve the minimal surface condition in \emph{two-dimensional} space
$(x^5, x^6)$, so the resulting surface is always a straight line (this
answer also follows from supersymmetry constraints). D5 branes are
represented by horizontal lines and called $(0,1)$ branes, and NS5
branes are vertical lines, or $(1,0)$ in this setting. When $(0,1)$
and $(1,0)$ branes merge one has to balance the tension, so the
resulting brane is can be neither vertical, nor horizontal. In fact
the tensions of $(0,1)$ and $(1,0)$ branes can be made the same by a
suitable choice of $x^5$ and $x^6$ scales, and we will assume this
choice throughout our discussion. The bound state of $(0,1)$ and
$(1,0)$ branes has unit slope, and is called the $(1,1)$ brane. All
other bound states can be obtained in a similar fashion and the whole
web of branes in the $(x^5, x^6)$ plane is called the Type IIB
$(p,q)$-\emph{brane web}~\cite{Aharony:1997bh}. The example of a brane
web is depicted in Fig.~\ref{fig:2}. Notice that the angles of the
branes are related to their charges --- which are conserved in any
brane merger. The moduli of the gauge theory correspond to the lengths
of the edges, which we denote by $Q_i$. More concretely, in the
example shown in Fig.~\ref{fig:2} $Q_{m,i}$ correspond to the gauge
theory masses, $Q_{F,i}$ encode the vacuum moduli and $Q_{B,i}$ are
related to the couplings $\Lambda_i$. Notice that in this picture the
vacuum averages, masses and coupling constants all appear on the same
footing. Moreover the $S$-duality of the type IIB theory turns $(1,0)$
brane into a $(0,1)$ brane, thus eliminating the asymmetry between the
vertical and horizontal directions.

Rotation of the brane web, or $S$-duality in Type IIB language, which
is obviously a symmetry of the theory, corresponds to a nontrivial
duality between 5d gauge theories, corresponding to the web. Indeed,
for $M$ vertical branes intersecting with $N$ horizontal branes one
has the $U(N)^{M-1}$ linear quiver, while for a rotated web it should
be $U(M)^{N-1}$. The relation between two gauge theories is given by
the \emph{spectral duality}. One can understand this duality as the
map on the space of BPS states of the theories, taking the $W$-bosons
of one theory into the instantons of the other and vice versa. The
masses of the BPS states are given by $M_{BPS} = |Z| \sim a_i n_i +
\Lambda_i k_i + \ldots$, where $n_i$ are $U(1)^N$ charges and $k_i$
are instanton numbers with respect to $U(N)^M$. From this expression
we see that the spectral duality maps Coulomb moduli of one theory to
couplings of the dual theory. This is also evident from the brane web,
since vertical and horizontal distances correspond to Coulomb moduli
and couplings respectively.

The integrable system corresponding to the 5d $U(N)^{M-1}$ linear
quiver gauge theory is the periodic $U_q(\mathfrak{gl}_M)$ XXZ spin
chain with $N$ spins \cite{GGM1}. As in the XXX case, masses and couplings of the
gauge theory are written in terms of Casimir operators and twist
matrix of the spin chain. Notice that spectral duality is also a
nontrivial duality of the two spin chains: the $U_q(\mathfrak{gl}_M)$
chain with $N$ spins is mapped onto the $U_q(\mathfrak{gl}_N)$ chain
with $M$ spins \cite{spedu}. The spectral curves of the two systems coincide, while
the Hamiltonians are expressed through each other and the parameters
of the systems.

Again, as in the Type IIA case, considered above, to obtain the
non-perturbative description of the $5d$ gauge theory, one has to lift
the $(p,q)$-web to M-theory. We start from Type IIB theory on
$\mathbb{R}^9 \times S^1_{R_4}$, which is equivalent to M-theory on
$\mathbb{R}^9 \times \mathbb{T}^2$ with $\mathrm{Vol}(\mathbb{T}^2) =
R_4$. A $(p,q)$-brane of Type IIB is M5 brane wrapping the
$(p,q)$-cycle on $\mathbb{T}^2$. By a chain of dualities one can
transform this brane picture into a purely geometric background
without any branes, namely the toric Calabi-Yau
threefold~\cite{Leung:1997tw}, so that the whole setup looks like
$\mathbb{R}^4 \times \mathrm{CY}_3 \times S^1$. $\mathrm{CY}_3$ is a
$\mathbb{T}^3$-fibration degenerating over the edges and vertices of
the \emph{toric diagram} $\mathcal{T} \in \mathbb{R}^3$, which
completely determines the geometry and is identified with the brane
web from Type IIB. Finite edges of the diagram represent the
two-spheres inside the CY, and, instead of the lengths of brane
segments, the relevant parameters are K\"ahler moduli of these
spheres, which we also denote by $Q_i$. Notice that not all the edge
lengths are independent: one needs to impose constraints, so that all
the cycles on the diagram are indeed closed. In particular, each cycle
imposes two constraints, corresponding to horizontal and vertical
projections. On Fig.~\ref{fig:2} we label only the independent
lengths.

\begin{figure}[h]
  \centering
\includegraphics[width=4.5cm]{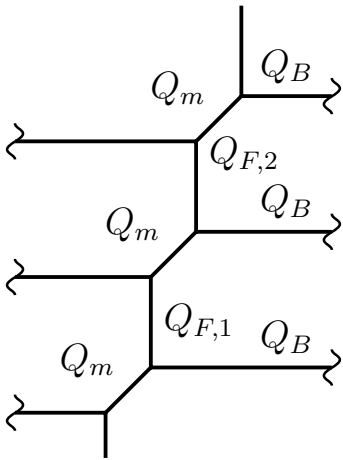}
\caption{Type IIB $(p,q)$-brane web, or CY toric diagram,
  corresponding to 5d $U(3)$ gauge theory with adjoint
  multiplet. Notice the cyclic identification of the lines on the left
  and right.}
  \label{fig:20}
\end{figure}

Toric or web diagrams can be compactified in the same way as the type
IIA brane diagram. Again, the edges of the picture become identified
and some of the branes wrap the resulting cycle, as shown on
Fig.~\ref{fig:20}, resulting in a $5d$ gauge theory with adjoint (or
more generally necklace quiver of bifundamental) matter. Moreover,
since there is no difference between the horizontal and vertical
directions in this picture, one can consider the vertical
compactification as well (see Fig.~\ref{fig:20}). The vertical
compactification corresponds to the 6d gauge theory with fundamental
matter. The important point here is that since the vertical direction
is compact, one cannot send the semi-infinite D5 branes up or down
indefinitely, and only $U(N)$ gauge theories with exactly $2N$
fundamental multiplets are allowed. Also, there is a constraint that
the sum of all the masses is equal to the sum of all the vacuum
moduli. In the language of the gauge theory this constraint can be
understood as the anomaly cancellation condition for large gauge
transformations.

Notice that spectral duality still acts on the compactified
diagram. However, the dual gauge theories are now of different
space-time dimension: one is the 5d $U(N)^M$ necklace quiver theory
and the other is the 6d $U(M)^{N-1}$ linear quiver. The BPS states of
the two theories are identified in the same way, as for the 5d linear
quivers discussed above.

In the language of integrable systems spectral duality gives a new
nontrivial identification between different \emph{elliptic}
systems. More concretely, 5d $U(N)^M$ necklace quiver theory
corresponds to multipoint generalization of elliptic Ruijsenaars
system, while 6d $U(M)^{N-1}$ linear quiver is described by the XYZ
spin chain. The two gauge theories are dual, and thus the integrable
systems are also dual, in the same way as the XXZ spin chains in the
example discussed above. We conclude that the \emph{XYZ spin chain is
  spectral dual to the multipoint elliptic Ruijsenaars model.}

Naturally one can also compactify \emph{both} horizontal and vertical
directions of the toric diagram, and obtain the 6d necklace quiver
theory. This theory seems in many ways special. In particular, the
corresponding Seiberg-Witten integrable system is of double elliptic
type, i.e.\ the Hamiltonians depend elliptically both on the
coordinates and momenta. This system is currently a subject of intense
investigation~\cite{Dell}. We only mention here that, since the doubly
compactified toric diagram can be rotated to get another doubly
compactified diagram, \emph{different double elliptic systems should
  be connected by the spectral duality}.

\subsection{From pictures to formulas}

For concrete computations, we use the relation between M-theory on
$\mathrm{CY}_3 \times \mathbb{R}^4 \times S^1$ and (refined)
topological strings on $\mathrm{CY}_3$, where the (refined)
topological vertex technique can be used. Let us describe the recipe
to obtain the partition function from the toric diagram.

To each 3-valent vertex of the toric diagram one assigns a 3-index
object $C_{ABC}(q,t)$, where $A$, $B$ and $C$ are Young diagrams,
residing on the three edges adjacent to the vertex, and the two
parameters of the vertex, $q=e^{-\epsilon_2 R_5}$,
$t=q^{\beta}=e^{\epsilon_1 R_5}$, are related to $\Omega$-deformation
parameters $(\epsilon_1, \epsilon_2)$ of the corresponding 5d gauge
theory on $\mathbb{R}^4 \times S^1_{R_5}$. In topological strings $q$
is also the exponentiated string coupling $q=e^{-g_s}$, so that $R_5$
can also be understood as the radius of the M-theory circle. The
ordering of indices in the refined topological vertex $C_{ABC}(q,t)$
depends on the extra labels of the adjacent edges. For toric
manifolds, which we consider, each vertex has one edge with preferred
direction (marked with double stroke) and two other labelled by $t$
and $q$, see \cite{IKV}:
\begin{multline}
  \label{eq:24}
  C_{ABC}(t,q) =\quad \parbox{2cm}{
    \includegraphics[width=2cm]{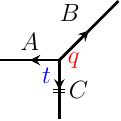}} \quad = q^{\frac{||B||^2 + ||C||^2}{2}}
  t^{-\frac{||B^{\mathrm{T}}||^2 + ||C^{\mathrm{T}}||^2}{2}} M^{(q,t)}_C
  \left(t^{-\rho}\right) \times\\
  \times\sum_D \left( \frac{q}{t}
  \right)^{\frac{|D|+|A|-|B|}{2}}  \chi_{A^{\mathrm{T}}/D}
  \left(q^{-C} t^{-\rho}\right) \chi_{B/D}\left(t^{-C^{\mathrm{T}}}
    q^{-\rho} \right),
\end{multline}

To each internal edge of the toric diagram one assigns
the (complexified) K\"ahler parameter $Q$ of the corresponding
two-cycle in the CY. If the two vertices are joined by an edge, then
the corresponding indices are contracted with the ``propagator''
$G^{AB}(Q,q,t)$:
\begin{gather}
  \label{eq:25}
  G^{AB}(Q,t,q) =\quad \parbox{2cm}{
    \includegraphics[width=2cm]{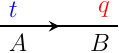}} \quad =
  \delta_{AB} Q^{|A|} \left( (-1)^{|A|}
    q^{\frac{||A^{\mathrm{T}}||^2+|A|}{2}}
    t^{-\frac{||A||^2+|A|}{2}} \right)^n,\\
  \widetilde{G}^{AB}(Q,t,q) =\quad \parbox{2cm}{
    \includegraphics[width=2cm]{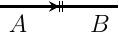}} \quad =
  \delta_{AB} Q^{|A|} \left( (-1)^{|A|}
    t^{\frac{||A^{\mathrm{T}}||^2}{2}}
    q^{-\frac{||A||^2}{2}} \right)^n,\label{eq:26}
\end{gather}
where $n$ is the framing factor, depending on the relative orientation
of the adjacent edges. Notice that the $t$ and $q$ marks should be
\emph{different} on the two sides of the propagator. The total closed
string partition function corresponding to a diagram is given by the
appropriate contraction of the propagators with the vertices. One can
also leave some of the sums over the diagrams unevaluated, to obtain
the open string amplitude. In this case, some of the external edges
will contain the Young diagram labels, corresponding to the open
string boundary conditions on the Lagrangian brane intersecting the
corresponding leg of the diagram.

\begin{figure}[h]
  \centering
  \includegraphics[width=5cm]{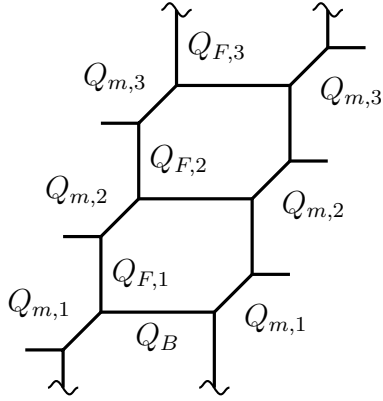}
  \caption{Type IIB $(p,q)$-brane web corresponding to 6d $U(3)$ gauge
    theory with 6 fundamental multiplets. The compactification radius
    is given by $e^{-\epsilon_2 R_6} = \prod_{a=1}^3 Q_{m,a}
    Q_{F,a}$.}
  \label{fig:40}
\end{figure}

Compactification of the toric diagram introduces a new parameter ---
which can be though as either the coupling constant $\tau$ of the
theory with adjoint hypermultiplet (similarly to Fig.~\ref{fig:10},
where it was given by $\Lambda$), or as the compactification radius
$R_6$ of the 6d theory. This parameter is encoded in the K\"ahler
parameter of the corresponding edge. For example, in Fig.~\ref{fig:20}
the (exponentiated) coupling constant is given by $Q_{\tau} = Q_B
Q_m$. Notice that in the CY construction of gauge theories there is
essentially no difference between the 5d theory with adjoint matter
(e.g.\ in Fig.~\ref{fig:20}) or a 6d theory with fundamental matter
(as e.g.\ in Fig.~\ref{fig:40}) --- which is just another
manifestation of the spectral duality.

We also observe a new curios feature of the compactified diagrams \cite{MMZ}. If
we consider the diagram corresponding to $U(1)$ gauge theory with
adjoint matter, we get extra poles in the corresponding DF
integral. Encircling these poles we arrive at the 6d $(1,1)$ gauge
theory on $\mathbb{R}^6_{q_1, q_2, q_3}$ with three equivariant
parameters. Two of these parameters are the equivariant parameters of
the original gauge theory, while the third one is given by the mass of
the adjoint multiplet. The partition function of this theory on
$\mathbb{C}^3 = \mathbb{R}^6$ computes the index topological
vertex~\cite{Nekrasov:2014nea} with external diagrams all empty. One can hope
that one more compactification of the brane diagram will give index
vertex with nonzero external legs.

\subsection{A primer: 5d $SU(2)$ theory and $q$-Virasoro 4-point conformal block}

The 4-point conformal block~\cite{CB} provided by the
Dotsenko-Fateev (conformal) matrix model~\cite{FDagt,DV,MMSS5} and AGT related
\cite{AGT} to the $N_f=4$ $SU(2)$ SUSY gauge theories~\cite{SW,Nekrasov:2002qd},
is associated with the square brane diagram~\cite{Gaiotto}.

The K\"ahler parameters of the diagram are related to gauge theory
parameters as follows\footnote{For simplicity we write the relations
  for $t=q$, i.e.\ $\beta=1$.}:
\begin{equation}
  \label{eq:45}
  Q_F = q^{2a},\qquad  Q^{\pm}_1 = q^{-m_1^{\pm} - a}, \qquad Q^{-}_2 =
  q^{m_2^{\pm} - a},\qquad  Q_B = \Lambda q^{2a}
\end{equation}
Here $a$ is the Coulomb modulus of the gauge theory $m_{1,2}^{\pm}$
are the masses of the four fundamental hypermultiplets and $\Lambda$
is the exponentiated complexified coupling constant.

\begin{figure}[h]
  \centering
  \includegraphics[width=6cm]{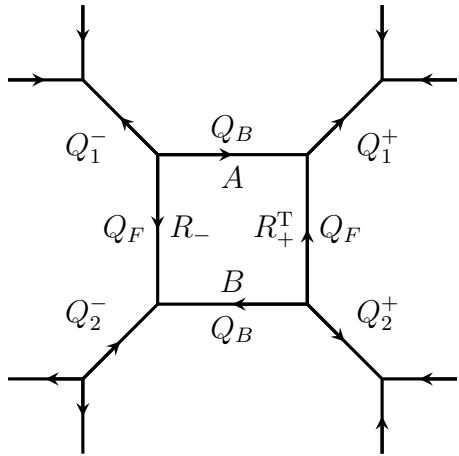}
  \caption{Toric diagram corresponding to the four-point $q$-deformed
    conformal block and $5d$ $U(2)$ gauge theory with four
    fundamentals.}
  \label{fig:33}
\end{figure}

\noindent
The DF representation reads
\begin{equation}
  \mathcal{B}_4^{\mathrm{sphere}}(\Delta_{\mathrm{int}}=a,\vec\Delta_{\mathrm{ext}},c = 1-6(b^2
  - b^{-2}),z=\Lambda) = \int \prod_{i=1}^Ndx_i\, \Delta^{2\beta}(x)
  V_{\alpha_0}(0,x) V_{\alpha_\Lambda}(\Lambda,x) V_{\alpha_1}(1,x)
\end{equation}
where $\Delta(x)$ is the Vandermonde determinant and $V_{\alpha} (z,x)
= \prod_{i=1}^N (z - x_i)^{b \alpha}$. On the other hand the block
should be given by a sum over a quadruple of partitions --- each
corresponding to internal edge of the diagram.

The hidden quadruple symmetry is revealed if Yang-Mills theory is
lifted to 5 dimensions \cite{Nek5,Braden} and ordinary Dotsenko-Fateev
integrals are substituted \cite{MMSS5} by Jackson sums
\begin{equation}
  \int_0^1 f(x) dx \longrightarrow (1-q) \sum_n q^n f(q^n)
  \label{eq:47}
\end{equation}
with discretization parameter $q=e^{-\epsilon_2 R_5}$. The Vandermonde
determinants in the measure are replaced by their $(q,t)$-deformation:
\begin{equation}
  \Delta^{(2\beta)}(x) \longrightarrow \prod_{m=1}^\beta \prod_{i,j=1}^N (x_i-q^{m-1} x_j)
  = \prod_{i,j=1}^N \prod_{m=0}^\infty \frac{x_i-q^m x_j}{x_i-q^mt x_j}
\end{equation}
with $t=q^\beta$. All the quantities in matrix model are analytically
continued from integer values of $N$ and $\beta$, what is made
unambiguously due to Selberg nature of the integrals \cite{MMMselb}.
This quadruple decomposition is recently presented in some detail in
\cite{Morozov:2015xya} based on the number of previous developments
\cite{MMMselb,MMSha,prev1,DV,prev2,MMSS5,genMD}, see sec.~\ref{sec:virasoro} below.  The
group theory symmetry behind the whole picture \cite{GKMMM} is encoded
in the 2-site $U_q(\mathfrak{gl}_2)$ XXZ spin chain integrable system
\cite{GGM2,MirMar}  (reduced to XXX in $4d$, when $q=1$ \cite{XXX}).

As evident from Fig.~\ref{fig:33}, the diagram is very symmetric. Some
symmetry is simple consequence of the symmetries in the matter content
of the corresponding gauge theory, e.g.\ reflections along the
horizontal and vertical lines correspond to renaming of the
fundamental hypermultiplets $m_1^{\pm} \leftrightarrow m_2^{\pm}$ and
charge conjugation respectively. Also, in the CFT language these
symmetries are related to the $z \to z^{-1}$ symmetry and renumbering
of the primary fields in the conformal block.

However, reflection along the \emph{diagonal} (or, equivalently
rotation by $\frac{\pi}{2}$) is not evident neither in the gauge
theory, nor in the CFT. In fact, it corresponds to the spectral
duality between two $U(2)$ gauge theories, such that Coulomb modulus
$q^{2a}$ is interchanged with the coupling constant
$\Lambda$\footnote{In~\cite{enhanced} this symmetries were embedded in
  the Weyl group of $E_5$ --- the global symmetry group of the
  interacting SCFT obtained from this setup as the UV fixed
  point.}. For the conformal blocks it is the duality between two
four-point conformal blocks of the $q$-Virasoro algebra, in which the
cross-ratio of positions is interchanged with the momentum of the
intermediate field.

Let us describe this peculiar duality in more detail. Conformal block
can be decomposed into a sum over intermediate states, and for certain
very specific choice of basis --- the basis of generalized Macdonald
polynomials --- labelled by pairs of Young diagrams, the resulting
decomposition reproduces the ``vertical cut'' sum over $A$ and $B$
diagrams in Fig.~\ref{fig:33}. Moreover, the decomposition is also
equal to the Nekrasov partition function and thus we get the following
equality between the topological string partition function, conformal
block and Nekrasov partition function:
\begin{multline}
  \label{eq:46}
  \underline{Z_{\text{top string}}} = \sum_{A,B} \Lambda
  ^{|A|+|B|}\langle V_{\alpha_0}(0) V_{\alpha_{\Lambda}} (1) | A,B,
  \Delta_{\mathrm{int}} \rangle \langle A,B, \Delta_{\mathrm{int}} |
  V_{\alpha_1}(1) V_{\alpha_{\infty}} (\infty) \rangle =
  \underline{\mathcal{B}_4^{\mathrm{sphere}} (\Delta_{\mathrm{ext}},
    \Delta_{\mathrm{int}}, \Lambda)}=\\
  =\sum_{A,B}
  \frac{z_{\mathrm{fund}}(a,m_i^{+},A,B)z_{\mathrm{fund}}(a,m_i^{-},A,B)}{z_{\mathrm{vect}}(a,A,B)}
  = \underline{Z_{\mathrm{Nek}}(a,m_i^{\pm},\Lambda)}
\end{multline}
Expressing the $q$-deformed conformal block $\mathcal{B}_4$ in the DF
form we get a discrete sum\footnote{There are two approaches to
  $q$-deformation of the DF and Selberg type integrals. One uses
  Jackson $q$-integrals (as in~\eqref{eq:47}), while the other employs
  contour integration and pole counting (see
  e.g.~\cite{Aganagic:2013tta,Sulk}). Both approaches are equivalent and we
  will in fact switch between them freely depending on which one is
  more convenient at the moment.}, which also turns out to be just the
``horizontal cut'' sum over $R_{\pm}$ in Fig.~\ref{fig:33}. Naturally,
since the diagram is symmetric, this sum is also a Nekrasov
decomposition of a certain gauge theory, though it is different from
the one featuring in Eq.~\eqref{eq:46} --- it is its spectral dual. We
have:
\begin{multline}
  \label{eq:48}
  \underline{Z_{\text{top string}}} = \sum_{R_{+}, R_{-}}
  \left. \left[ \Delta^{2\beta}(x) V_{\alpha_0}(0,x)
      V_{\alpha_\Lambda}(\Lambda,x) V_{\alpha_1}(1,x)
    \right]\right|_{x_{\pm ,i} = q^{R_{\pm ,i} + 1} t^{N_{\pm-i}}} =
  \underline{\mathcal{B}_4^{\mathrm{shpere}} (\Delta_{\mathrm{ext}},
    \Delta_{\mathrm{int}}, \Lambda)} = \\
  = \sum_{R_{+}, R_{-}} \Lambda_{\mathrm{d}}^{|R_{+}|+|R_{-}|}
  \frac{z_{\mathrm{fund}}(a_{\mathrm{d}},m_{\mathrm{d},i}^{+},A,B)z_{\mathrm{fund}}(a_{\mathrm{d}},m_{\mathrm{d},i}^{-},A,B)}{z_{\mathrm{vect}}(a_{\mathrm{d}},A,B)}
  =
  \underline{Z_{\mathrm{Nek}}(a_{\mathrm{d}},m_{\mathrm{d},i}^{\pm},\Lambda_{\mathrm{d}})}
\end{multline}
In the last line the parameters of the Nekrasov function are dual, in
particular the Coulomb moduli and coupling constant are exchanged.

Spectral duality acts on the topological vertices trivially only for
$t=q$. In the \emph{refined} case, $t\neq q$, and the vertices
transform nontrivially under the rotation of the diagram, since the
preferred direction also changes (see~\eqref{eq:24}). However, in this
case one can understand the \emph{algebraic} meaning of the
duality.

Instead of considering rotation of the brane diagram, one can consider
the rotation of the preferred direction. Different choices of the
preferred direction correspond to different choices of basis in the
tensor product of Fock modules corresponding to the legs of the
diagram. For preferred direction along the legs the basis
$|A,B,\Delta\rangle$ is the one appearing in the
decomposition~\eqref{eq:46}. For an orthogonal choice of the preferred
direction, the basis is \emph{factorized} $|A \rangle \otimes |B
\rangle$. The change of basis from $|A,B,\Delta\rangle$ to the
factorized basis $|A \rangle \otimes |B \rangle$ is performed using
the generalized Kostka functions, described
in~\cite{Morozov:2015xya}. We will return to this point when we
describe the action of spectral duality for the compactified diagram.

\section{DF measures from toric diagram}
\label{sec:measures}

How do the DF measures for different conformal blocks arise from the toric
diagrams? We show how to glue them from the elementary building block
  \begin{multline}
    Z \left( \left.
  \begin{smallmatrix}
     & P & \\
     A&  & B\\
     & R &
  \end{smallmatrix}
\right| Q, q,t \right) \quad =
\quad \parbox{2.8cm}{\includegraphics[width=2.8cm]{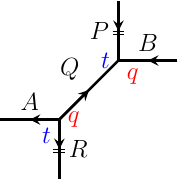}}=Z_{\varnothing}(Q)
q^{\frac{||R||^2 - ||P||^2}{2}} t^{\frac{||P^{\mathrm{T}}||^2 -
    ||R^{\mathrm{T}}||^2}{2}} \left( \frac{q}{t}
\right)^{\frac{|A| - |B|}{2}}\times \\
\times M_R^{(q,t)}(t^{-\rho}) M_{P^{\mathrm{T}}}^{(t,q)} (q^{-\rho})
\prod_{i,j \geq 1} \prod_{k\geq 0} \frac{\left( 1 - q^k
    \sqrt{\frac{q}{t}} Q q^{R_i-P_j} t^{j - i+1}\right) \left( 1 - q^k
    \sqrt{\frac{q}{t}} Q t^{j - i} \right) }{\left( 1 - q^k
    \sqrt{\frac{q}{t}} Q q^{R_i-P_j} t^{j - i}\right) \left( 1 - q^k
    \sqrt{\frac{q}{t}}
    Q t^{j - i +1} \right) } \times\\
\times (-Q)^{|A|} \sum_C \chi_{A/C} \left( p_n \left(
    \sqrt{\frac{q}{t}}  q^{-P} t^{-\rho}
  \right) - p_n (Q^{-1}
  q^{-R} t^{-\rho})\right) \chi_{B/C} \left( \frac{1 - t^n}{1 - q^n}
  \left[p_n \left( Q q^R t^{\rho} \right) - p_n ( \sqrt{\frac{q}{t}}q^P t^{\rho})\right]  \right)=\\
= \mathcal{N} M_R^{(q,t)}(t^{-\rho}) M_{P^{\mathrm{T}}}^{(t,q)}(q^{-\rho})
G_{RP}^{(q,t)} \left( \sqrt{\frac{q}{t}} Q \right) \times\\
\times \sum_C (-Q)^{|C|} \chi_{A^{\mathrm{T}}/C^{\mathrm{T}}}
\left(p_n (t^{-\rho} q^{-R}) - p_n \left( \sqrt{\frac{q}{t}} Q
    t^{-\rho} q^{-P} \right)\right) \chi_{B/C} \left( p_n (q^{-\rho}
  t^{-P^{\mathrm{T}}}) - p_n \left( \sqrt{\frac{t}{q}} Q q^{-\rho}
    t^{-R^{\mathrm{T}}} \right) \right) \label{eq:13}
  \end{multline}
where $Z_{\varnothing}(Q) = Z \left( \left.
  \begin{smallmatrix}
     & \varnothing & \\
     \varnothing&  & \varnothing\\
     & \varnothing &
  \end{smallmatrix}
\right| Q, q,t \right)$ and $\mathcal{N}$ collects both the factors
independent of the external diagrams $A$, $B$, $R$ and $P$ and framing
factors which cancel when several amplitudes are glued together. When
gluing two four-point amplitudes we use the identity~\eqref{eq:3}.

It is possible to contract the amplitudes~\eqref{eq:13} in many
different ways, and in the next several sections we will investigate
all these possibilities. First of all, there are planar rectangular
webs. They correspond to 5d linear quiver gauge theories with gauge
groups $U(N)^M$. One can compactify the rectangle by identifying the
opposite edges. Then, one gets either 6d linear quiver gauge theory or
5d necklace quiver, depending on the orientation of the identified
edges. The two descriptions are spectral dual to each other, since
they turn into one another by $\frac{\pi}{2}$ rotation of the brane
diagram. Identifying both vertical and horizontal edges of the
rectangle give the 6d necklace quiver gauge theory. These theories are
spectral self-dual in a sense that the 6d theory with gauge group
$U(N)^M$ is dual to the theory with gauge group $U(M)^N$. This
generalizes the familiar result for 5d theories~\cite{Bao}.

The classification of different compactifications/deformations and the
respective measures is summarized in the table:
\begin{center}
  \begin{tabular}{c|c|c}
    & Sphere & Torus\\
    \hline
    algebra&Selberg&Elliptic Selberg\\
  $q$-deformed algebra & $q$-Selberg& Elliptic $q$-Selberg\\
  doubly-deformed algebra & Affine Selberg& Elliptic affine Selberg\\
\end{tabular}
\end{center}

In each case there can be either Virasoro or $W_M$-algebra conformal
block, corresponding to $A_1$ or $A_{M-1}$ Selberg measure. There can
also be multiple vertex operator insertions in the integral. The
number of these insertions and the rank of the $W_M$ algebra are
related by the spectral duality.

\subsection{$q$-deformed spherical block}
\label{sec:q-deformed-spherical}

The block is given by multiple contour integrals, each corresponding
to a primary field insertion. These integrals can be evaluated by
residues~\cite{Aganagic:2013tta} and the poles are enumerated by Young
diagrams, so that the resulting expressions coincide with the
combinations of four-point topological string amplitudes. The
alternative approach is to write down the sum instead of an integral
from the very beginning. This sum is then called the Jackson
$q-$integral and is defined as
\begin{equation}
  \label{eq:15}
  \int_0^a d_qx f(x) = (1-q) \sum_{n \geq 0} q^n a f(q^n a)
\end{equation}

\subsubsection{Virasoro conformal block}
\label{sec:virasoro}
In this case the elementary building block of the DF integral is the
$q$-Selberg average:
\begin{equation}
  \label{eq:10}
  \langle f(x) \rangle_{q,t} =\int_0^1 d_q^N x\, \Delta^{(q,t)}(x) \prod_{i=1}^N x_i^u
  \prod_{k = 0}^{v-1} (1 - q^k x_i) f(x_i),
\end{equation}
where the essential part of the measure is the $(q,t)$-Vandermonde
\begin{equation}
  \label{eq:31}
  \Delta^{(q,t)}(x) = \prod_{i \neq j} \prod_{k \geq 0}\frac{1
      - q^k \frac{x_i}{x_j}}{1
      - t q^k \frac{x_i}{x_j}} = \prod_{i \neq j} \prod_{k = 0}^{\beta-1} \left(1
    - q^k \frac{x_i}{x_j}\right).
\end{equation}

It corresponds to the following combination of four-point amplitudes:
\begin{multline}
  \label{eq:9}
  \parbox{4cm}{\includegraphics[width=4cm]{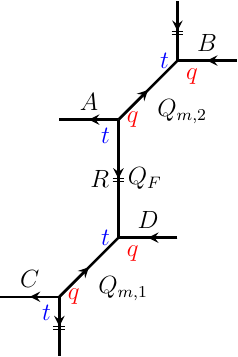}}\quad
  =  \mathcal{N}  (-Q_{m,1})^{|A|} (-Q_{m,2})^{|C|}  \sum_R \left( \sqrt{\frac{t}{q}} Q_F \right)^{|R|} \times\\
  \times \prod_{k = 0}^{\beta-1}\prod_{i \neq j \geq 1} \frac{ \left(1
      - q^k \frac{x_{R,i}}{x_{R,j}}\right) \left(1 - q^k Q_{m,1}
      Q_{m,2} \frac{t^{N-i}}{t^{N-j}}\right) }{\left( 1 - q^k
      \sqrt{\frac{q}{t}} Q_{m,1} \frac{t^{N-i}}{x_{R,j}} \right)
    \left( 1 - q^k \sqrt{\frac{q}{t}} Q_{m,2} \frac{x_{R,i}}{t^{N-j}}
    \right)}\times \\
  \times \sum_{E,F} \chi_{A/E} \left( p_n \left( \sqrt{\frac{q}{t}}
      t^{-\rho} \right) -p_n \left( Q_{m,2}^{-1} t^{N + \frac{1}{2}}
      x_R^{-1} \right) \right) \chi_{B/E} \left( \frac{1 - t^n}{1 -
      q^n} \left( p_n \left( Q_{m,2} t^{-\frac{1}{2} - N} x_R \right)
      - p_n \left( \sqrt{\frac{q}{t}} t^{\rho}
      \right) \right) \right)\times\\
  \times \chi_{C/F} \left( p_n \left( \sqrt{\frac{q}{t}} t^{N +
        \frac{1}{2}} x_R^{-1} \right) -p_n \left( Q_{m,1}^{-1}
      t^{-\rho} \right) \right) \chi_{D/F} \left( \frac{1 - t^n}{1 -
      q^n} \left( p_n \left( Q_{m,1} t^{\rho} \right) - p_n \left(
        \sqrt{\frac{q}{t}} t^{-\frac{1}{2} - N} x_R \right) \right)
  \right)
\end{multline}
where $x_{R,i} = q^{R_i + 1}t^{N-i}$ and $\mathcal{N}$ is a
normalization constant independent of the diagrams $A$, $B$, $C$,
$D$. To get the correct $q$-Selberg measure we have to set
$\sqrt{\frac{q}{t}} Q_{m,1} = q ^{-v}t^{-N}$, $\sqrt{\frac{q}{t}}
Q_{m,2} = t^N$, and $Q_F = q^{u+v+\frac12} t^{-\frac12}$. Let us
understand the structure of the topological string amplitude. The
$(q,t)$-Vandermonde appears in the second line. The product over $i$
and $j$ is in fact finite for the discrete choice of the K\"ahler
parameters we have made, since most terms in the numerator and the
denominator cancel with each other. After the dust settles we get the
residues of the $q$-Selberg measure:
\begin{multline}
  \label{eq:14}
  \left( \sqrt{\frac{t}{q}} Q_F \right)^{|R|} \prod_{k =
    0}^{\beta-1}\prod_{i \neq j \geq 1} \frac{ \left(1 - q^k
      \frac{x_{R,i}}{x_{R,j}}\right) \left(1 - q^k Q_{m,1} Q_{m,2}
      \frac{t^{N-i}}{t^{N-j}}\right) }{\left( 1 - q^k
      \sqrt{\frac{q}{t}} Q_{m,1} \frac{t^{N-i}}{x_{R,j}} \right)
    \left( 1 - q^k \sqrt{\frac{q}{t}} Q_{m,2} \frac{x_{R,i}}{t^{N-j}}
    \right)} =\\
  =\frac{\Delta^{(q,t)}(x_R) \prod_{i=1}^N x_{R,i}^u
  \prod_{k=0}^{v-1} \left(1 - q^k x_{R,i} \right)}{\Delta^{(q,t)}(x_{\varnothing}) \prod_{i=1}^N x_{\varnothing,i}^u
  \prod_{k=0}^{v-1} \left(1 - q^k x_{\varnothing,i} \right)} =
  \frac{\mu_{A_1}(x_R|u,v,N,q,t)}{\mu_{A_1}(x_{\varnothing}|u,v,N,q,t)}.
\end{multline}
This exactly reproduces the original $q$-Selberg
measure~\eqref{eq:10}. The last two lines in Eq.~\eqref{eq:9} contain
Schur functions, which are being averaged with the $q$-Selberg
measure. They play the role of the function $f(x)$ in
Eq.~\eqref{eq:10}.

\subsubsection{$W_M$-algebra conformal block}
\label{sec:w_n}

The next logical step is to consider the $W_M$ conformal block. In our
formalism only special primary fields with weight proportional to a
single fundamental weight are allowed. This reproduces the case of AGT
correspondence, where these fields correspond to bifundamental field
insertions in the $SU(M)$ gauge theory partition function. The
relevant toric diagram for this block is the vertical strip geometry
obtained by gluing together $M$ copies of the four-point
amplitude~\eqref{eq:13} in the vertical direction.

\begin{figure}[h]
  \centering
  \includegraphics[width=5cm]{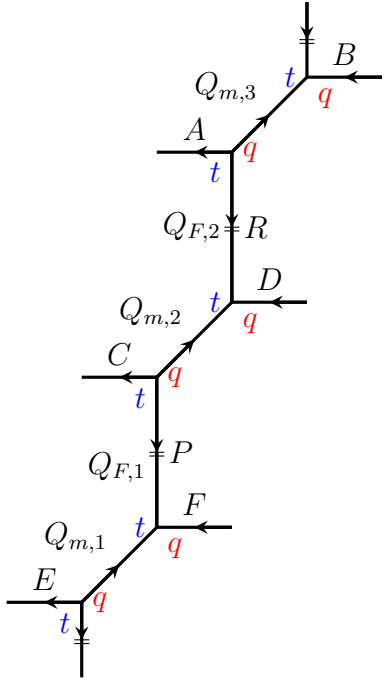}
  \caption{Three four-point amplitudes glued together vertically give
    the $A_2$ $q$-Selberg measure.}
\end{figure}

The K\"ahler parameters should be identified as follows:
\begin{gather}
  \label{eq:18}
  \sqrt{\frac{q}{t}} Q_{m,1} = q^{-v} t^{-N_1},\qquad
  \sqrt{\frac{q}{t}} Q_{m,a} = t^{N_{a-1} - N_a}\quad \text{for}\quad
  a = 2,\ldots, M-1,\qquad
  \sqrt{\frac{q}{t}} Q_{m,M} = t^{N_{M-1}},\\
  Q_{F,a} = q^{u_a + \delta_{a,1} v + \frac12} t^{-\frac12}.
\end{gather}
We skip the tedious technical details and give the final answer for
the $A_{M-1}$ $q$-Selberg measure:
\begin{equation}
  \label{eq:19}
  \mu_{A_{M-1}}(\vec{x}|\vec{u},v,\vec{N},q,t)
  = \Delta^{(q,t)}_{A_{M-1}} (x) \left( \prod_{a=1}^{M-1}
    \prod_{i=1}^{N_a} x_{a,i}^{u_a} \right) \prod_{i=1}^{N_1} \prod_{k=0}^{v-1}
  \left( 1 - q^k x_{1,i} \right),
\end{equation}
where essential part of the measure is the $A_{M-1}$ $q$-Vandermonde
determinant, which is given by
\begin{equation}
  \label{eq:12}
  \Delta^{(q,t)}_{A_{M-1}}(\vec{x}_{\vec{R}}) =
  \frac{\Delta^{(q,t)}(x_{R_1})\cdots
    \Delta^{(q,t)}(x_{R_{M-1}})}{\Delta^{(q,t)}(x_{R_1},
    x_{R_2})\cdots \Delta^{(q,t)}(x_{R_{M-2}}, x_{R_{M-1}})}
\end{equation}
where $x_{R_a,i} = q^{R_{a,i}+1} t^{N_a - i}$, $\Delta^{(q,t)}(x)$ is
given by Eq.~\eqref{eq:31} and
\begin{equation}
  \label{eq:32}
  \Delta^{(q,t)}(x,y) = \prod_{k\geq 0} \prod_{i=1}^{N_1}
  \prod_{j=1}^{N_2} \frac{1 - q^k \frac{x_i}{y_j}}{1 - t q^k \frac{x_i}{y_j}}
\end{equation}
Here $M-2$ denominators appear from $M$ four-point amplitudes
like~\eqref{eq:13} and $M-1$ numerators arise from the
identity~\eqref{eq:3} used to glue them together. Notice also that the
vertex operator contribution $\prod (1 - q^k x_{1,i})$ contains only
the variables $x_{1,i}$ from the first integration contour. This is
the general effect, seen in the $SU(N)$ versions of the AGT conjecture
--- the vertex operators, corresponding to Lagrangian quiver gauge
theories are not general $W_N$ primary fields, but have dimension
vector proportional to a single fundamental weight. We conclude that
our gluing procedure correctly reproduces both the gauge theory
results and the corresponding $W_N$ conformal blocks.

One can clearly sees the structure of the root system
$A_{N-1}$ in the measure~\eqref{eq:12}. Other classic root systems
$\Phi$ can be obtained from orbifolding the toric CY, so that in
general the $q$-Vandermonde is given by
\begin{equation}
  \label{eq:11}
  \Delta^{(q,t)}_{\Phi}(\vec{x}_{\vec{R}}) = \prod_{k=0}^{\beta-1} \left[\prod_{a<b}
    (\Delta^{(q,t)}(x_a, x_b))^{A_{ab}}  \prod_a\Delta^{(q,t)}(x_a) \right],
\end{equation}
where $A_{ab}$ is the Cartan matrix. The ADE case of this measure was
derived in~\cite{Aganagic:2015cta} and in~\cite{Kimura:2015rgi}.

\subsection{$q$-deformed torus block}
\label{sec:q-deformed-toric}
We next consider the four point amplitude with \emph{horizontal} ends
identified.
\begin{multline}
  \label{eq:21}
  \parbox{3cm}{\includegraphics[width=3cm]{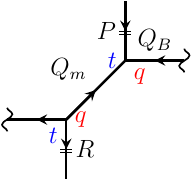}} =
  \mathcal{N} q^{\frac{||R||^2-||P||^2}{2}}
  t^{\frac{||P^{\mathrm{T}}||^2 - ||R^{\mathrm{T}}||^2}{2}}
  M_R^{(q,t)}(t^{-\rho}) M_{P^{\mathrm{T}}}^{(t,q)}(q^{-\rho})\times\\
  \times \prod_{k \geq 1} \frac{G_{RP}^{(q,t)} \left( Q_{\tau}^{k-1}
      \sqrt{\frac{q}{t}} Q_m \right) G_{PR}^{(q,t)} \left(
      Q_{\tau}^k \sqrt{\frac{q}{t}} Q_m^{-1} \right)}{G_{PP}^{(q,t)}
    \left( Q_{\tau}^k \frac{q}{t} \right) G_{RR}^{(q,t)} \left(
      Q_{\tau}^k \right)},
\end{multline}
where $Q_{\tau} = Q_m Q_B$ and the factor $\mathcal{N}$ is independent
of the diagrams $R$, $P$. It is possible to express some infinite
products in the r.h.s.\ of Eq.~\eqref{eq:21} in the form of
theta-functions (see e.g.~\cite{Haghighat:2013gba}), however we will
not be concerned with modular properties and thus will not need this
representation.

Let us describe the action of spectral duality for the compactified
toric diagram. It is again the $\frac{\pi}{2}$ rotation of the
diagram, and for the case of $t=q$ this is an explicit symmetry of the
topological string formalism. Using this formalism one can again see
that the poles in DF integral representation of the conformal block
correspond to AGT-like decomposition of the spectral dual
conformal block.

However, for $t\neq q$, refined vertex should be used. This
vertex~\eqref{eq:24} is not rotation symmetric, and thus spectral
duality requires a nontrivial change of basis of states of the
topological string, which we briefly explained in the
Introduction. This basis change from generalized Macdonald polynomials
to a factorized basis (of Schur functions) is given by the nontrivial
generalized Kostka functions. These matrix functions act in the tensor
product of several Fock spaces and, therefore, depend on several Young
diagrams. The number of diagrams is given by the number of parallel
edges in the toric diagram. The case of compactified toric diagram
corresponds to effectively infinite number of legs --- one can
understand the leg on a circle as an infinite array of ``mirror
images'' in the uncompactified space.

Unfortunately, in this case the generalized Macdonald
polynomials are in fact not yet known. Let us only notice that, unlike
the uncompactified case, here the problem is nontrivial even for a
single leg, i.e.\ a single Fock module, where the basis is labelled by
a single Young diagram (though, of course depends on an extra
parameter --- the compactification radius).

\subsubsection{Virasoro conformal block}
\label{sec:virasoro-1}

Gluing two blocks~\eqref{eq:21} together we get the $q$-deformed
Virasoro DF integral on torus, which is also known as the elliptic
version of the $A_1$ Selberg integral:
\begin{multline}
  \label{eq:22}
  \parbox{4cm}{\includegraphics[width=4cm]{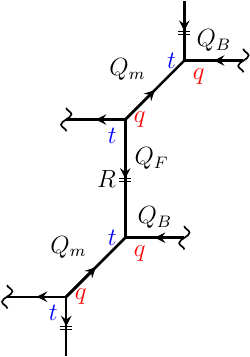}} = \mathcal{N} \sum_R \left( Q_F \sqrt{\frac{t}{q}} \right)^R
  \prod_{k \geq 1} \frac{G_{R \varnothing}^{(q,t)}\left(
      Q_{\tau}^{k-1} \sqrt{\frac{q}{t}} Q_m \right)  G_{\varnothing R}^{(q,t)}\left(
      Q_{\tau}^{k-1} \sqrt{\frac{q}{t}} Q_m \right)}{G_{RR}^{(q,t)}
    \left( Q_{\tau}^{k-1} \right) }\times\\
  \times \prod_{k \geq 1} \frac{G_{\varnothing R}^{(q,t)}\left(
      Q_{\tau}^k \sqrt{\frac{q}{t}} Q_m^{-1} \right) G_{R \varnothing}^{(q,t)}\left(
      Q_{\tau}^k \sqrt{\frac{q}{t}} Q_m^{-1} \right)}{G_{RR}^{(q,t)} \left(
      \frac{q}{t} Q_{\tau}^k \right)}
\end{multline}
where $Q_{\tau} = Q_m Q_B$. The sum over partitions in
Eq.~\eqref{eq:22} can be recast into the contour integral of the
\emph{elliptic} Selberg form:
\begin{equation}
  \label{eq:27}
  \oint d^Nx\, \Delta^{(q,t,Q_{\tau})}(x) V_u(0,x) V_N(1,x),
\end{equation}
where
\begin{gather}
  \label{eq:28}
  \Delta^{(q,t,Q_{\tau})}(x) = \prod_{k \geq 1} \prod_{l \geq 0}
  \prod_{i \neq j}^N
  \frac{\left( 1 - q^l Q_{\tau}^{k-1} \frac{x_i}{x_j} \right) \left( 1
      - q^l Q_{\tau}^k \frac{q}{t} \frac{x_i}{x_j} \right)}{\left( 1
      - q^l Q_{\tau}^{k-1} t \frac{x_i}{x_j} \right) \left( 1 - q^l
      Q_{\tau}^k q \frac{x_i}{x_j} \right)}, \qquad V_u(0,x) =
  \prod_{i=1}^N x_i^u\\
 V_v(z,x) = \prod_{i=1}^N \prod_{k \geq 1} \prod_{l \geq 0} \frac{\left(1 -
   q^l Q_{\tau}^{k-1} q^{-v} \frac{z}{x_i}\right)\left(1 -
   q^l Q_{\tau}^k q^{-v} \frac{q}{t} \frac{z}{x_i}\right)}{\left(1 -
   q^l Q_{\tau}^{k-1} \frac{z}{x_i}\right)\left(1 -
   q^l Q_{\tau}^k \frac{q}{t} \frac{z}{x_i}\right)}
\end{gather}
Here $t^N = \sqrt{\frac{q}{t}} Q_m$ and $q^u = Q_F
\sqrt{\frac{t}{q}}$. Notice the symmetry between $q$ and $Q_{\tau}$ in
the measure --- this follows from the equivalence between the two
compactified circles $S_{R_5}$ and $S_{R_6}$. This form of elliptic
Selberg integral was used in~\cite{6dAGT} to formulate the elliptic
version of the AGT correspondence. Notice that the dimension of the
vertex operator $V_N(1,x)$ is not an independent variable, but is
related to the number of integrations, and thus to the intermediate
dimension in the toric conformal block.

\subsubsection{$W_M$-algebra conformal block}
\label{sec:w_m-algebra}
The same procedure as in sec.~\ref{sec:virasoro-1} applies to the
$W$-algebra case. We glue several compactified pieces
like~\eqref{eq:21} together to obtain the toric diagram corresponding
to the 5d $U(M)$ gauge theory with adjoint multiplet.

The corresponding $A$ elliptic Selberg measure is given by
\begin{equation}
  \label{eq:29}
    \Delta^{(q,t,Q_{\tau})}_{A_{M-1}}(\vec{x}) =
    \frac{\Delta^{(q,t,Q_{\tau})}(x_1)\cdots
      \Delta^{(q,t,Q_{\tau})}(x_{M-1})}{\Delta^{(q,t,Q_{\tau})}(x_1, x_2)
      \cdots \Delta^{(q,t,Q_{\tau})}(x_{M-2}, x_{M-1})},
\end{equation}
where $\Delta^{q,t,Q_{\tau}}(x_a)$ is given by Eq.~\eqref{eq:28} and
\begin{equation}
  \label{eq:30}
  \Delta^{(q,t,Q_{\tau})}(x, y) = \prod_{k\geq 1}\prod_{i=1}^{N_1} \prod_{j=1}^{N_2} \frac{\left( 1 - q^l Q_{\tau}^{k-1} \frac{x_i}{y_j} \right) \left( 1
      - q^l Q_{\tau}^k \frac{q}{t} \frac{y_j}{x_i} \right)}{\left( 1
      - q^l Q_{\tau}^{k-1} t \frac{x_i}{y_j} \right) \left( 1 - q^l
      Q_{\tau}^k q \frac{y_j}{x_i} \right)}
\end{equation}

Of course, the measure for any root system $\Phi$ can be obtained in a
similar way:
\begin{equation}
  \label{eq:34}
  \Delta_{\Phi}^{(q,t,Q_{\tau})} (\vec{x}) = \left[\prod_{a=1}^r
  \Delta^{(q,t,Q_{\tau})}(x_a) \right] \prod_{a<b}
  \Delta^{(q,t,Q_{\tau})}(x_a, x_b)^{A_{ab}},
\end{equation}
where $A_{ab}$ is the Cartan matrix of the root system $\Phi$.

\subsection{Affine spherical block}
\label{sec:ellipt-spher-block}

This case provides the vertex operator corresponding to the 6d gauge
theory. It is very much analogous to the $W_M$ case considered above,
however the root system is now not finite, but \emph{affine}. The
resulting algebra is \emph{doubly} deformed --- by $q$ and by an extra
parameter $\tilde{t}$ related to the compactified sixth dimension. We
start from the simplest example of $U(1)$ gauge theory and then
generalize to higher rank groups. We also present an interesting
generalization of the affine integrals in
sec.~\ref{sec:u1-theory-with}.

\subsubsection{$\widehat{\mathfrak{u}}(1)$ measure}
\label{sec:widehatu1-q-selberg}
The simplest example of 6d theory is the $U(1)$ linear quiver. It is
built from the elementary block corresponding to the bifundamental
field. In the topological string language this corresponds to a four
point amplitude with vertical edges joined with each other.

\begin{figure}[h]
  \centering
  \includegraphics[width=3cm]{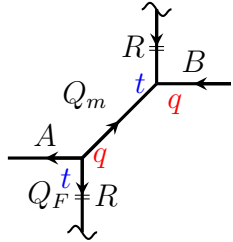}
  \caption{Four-point amplitude with vertical ends identified to
    obtain the $\widehat{\mathfrak{u}}_1$ $q$-Selberg measure.}
  \label{fig:u1hat}
\end{figure}

The new feature appearing in this setting is that the width of the
diagram featuring in the sum is not bounded, so that the number of
$x_i$ variables in the corresponding integral becomes infinite:
\begin{equation}
  \label{eq:16}
  \Delta^{(q,t,\tilde{t})}_{\widehat{\mathfrak{u}}_1} (x_R) 
    = \lim_{N\to \infty} \prod_{i\neq j}^N
  \prod_{k\geq 0} \frac{\left(1 - q^k
      \frac{x_{R,i}}{x_{R,j}}\right) \left(1 - t \tilde{t} q^k \frac{x_{R,i}}{x_{R,j}}\right) }{\left(1 - t q^k \frac{x_{R,i}}{x_{R,j}}\right) \left(1 - \tilde{t} q^k
        \frac{x_{R,i}}{x_{R,j}}\right)}
\end{equation}
where $\tilde{t} = \sqrt{\frac{q}{t}} Q_m$. Notice the symmetry
between $t$ and $\tilde{t}$, which is made explicit in the last
expression. This should be compared to the analogous symmetry between
$q$ and $Q_{\tau}$ in the \emph{toric} measure~\eqref{eq:28}. Of
course, the two cases are related by the spectral duality, which
partly explains the similarity. However, the exact relation between
the two measures is nontrivial, since the sum over $R$ in
Fig.~\ref{fig:u1hat} is over a different edge, than in
Eq.~\eqref{eq:22}. We will look more closely at the
measure~\eqref{eq:16} in sec.~\ref{sec:u1-theory-with}.

\subsubsection{$\widehat{A}_M$ measure}
\label{sec:widehata_m-measure}

\begin{figure}[h]
  \centering
  \includegraphics[width=5cm]{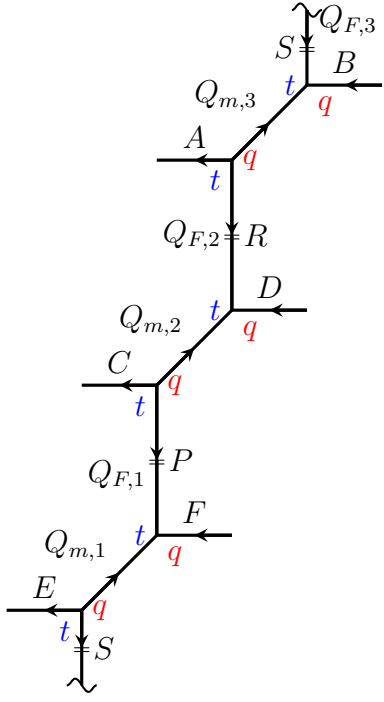}
  \caption{Three four-point amplitudes glued vertically in a cyclic
    fashion to obtain the $\widehat{A}_2$ $q$-Selberg measure.}
\end{figure}

The story is similar for several four-point amplitudes glued in a
cyclic fashion in the vertical direction. The corresponding
Vandermonde factor is
\begin{equation}
  \label{eq:20}
  \Delta^{(q,t,\tilde{t})}_{\widehat{A}_{M-1}} (\vec{x}_{\vec{R}}) =
  \lim_{N_a \to \infty}  \frac{\prod_{a=1}^M
    \Delta^{(q,t)} (x_{R_a})}{\Delta^{(q,t)}
    (\tilde{t} x_{R_1}, x_{R_M}) \prod_{a=1}^{M-1} \Delta^{(q,t)}
    (x_{R_{a+1}}, x_{R_a})}
\end{equation}
where differences between $N_a$ are encoded by the K\"ahler parameters
of the diagram: $t^{N_{a+1} - N_a} = \sqrt{\frac{q}{t}} Q_{m,a}$,
$a=1,\ldots, M-1$, and $ t^{N_1-N_M} \tilde{t} = \sqrt{\frac{q}{t}}
Q_{m,M}$. Notice that the product corresponding to the additional
\emph{affine} root is distinguished, and contains an extra parameter
$\tilde{t}$. Of course, one can rescale the variables $x_{R_a}$ so
that the extra parameter appears in any other product. The other
curious fact about the affine case is that the number of contours
(and, thus, the number of Young diagrams in the sum over poles) in the
$\widehat{A}_M$ measure is $M$, and not $M-1$ as in the $A_M$
case. This special point is closely related to the fact that the
bosonization of the \emph{elliptic} Virasoro algebra necessarily
involves free fields of \emph{two} sorts, whereas only \emph{one} free
field is needed in the rational and $q$-deformed cases.

\subsection{Affine torus block}
\label{sec:elliptic-toric-block}

We proceed to describe the most advanced generalization, i.e.\ the
double compactification of the diagram. It corresponds to the torus
conformal block for the $W$-algebra of the affine algebra (also
denoted by $W(\widehat{\mathfrak{g}})$).

\subsubsection{Torus $\widehat{\mathfrak{u}}_1$ measure}
\label{sec:wideh-meas}

\begin{figure}[h]
  \centering
  \includegraphics[width=3cm]{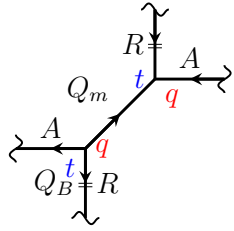}
  \caption{Four-point amplitude with both vertical and horizontal ends
    identified to obtain the $\widehat{\mathfrak{u}}_1$ elliptic
    $q$-Selberg measure.}
  \label{fig:u1hathat}
\end{figure}

\begin{equation}
  \label{eq:33}
  \Delta_{\widehat{\mathfrak{u}}_1}^{(q,t,\tilde{t},Q_{\tau})}(x) =
  \lim_{N \to \infty} \prod_{i \neq j} \prod_{k \geq 0}
  \prod_{l \geq 0} \frac{\left(1 - Q_{\tau}^k q^l
      \frac{x_i}{x_j}\right) \left(1 - Q_{\tau}^k q^l \frac{Q_{\tau} q}{t}
      \frac{x_i}{x_j}\right)}{\left(1 - Q_{\tau}^k q^l t
      \frac{x_i}{x_j}\right) \left(1 - Q_{\tau}^k q^l Q_{\tau} q
      \frac{x_i}{x_j}\right)} \frac{\left(1 - Q_{\tau}^k q^l t \tilde{t}
      \frac{x_i}{x_j}\right) \left(1 - Q_{\tau}^k q^l Q_{\tau} q \tilde{t}
      \frac{x_i}{x_j}\right)}{\left(1 - Q_{\tau}^k q^l
      \tilde{t} \frac{x_i}{x_j}\right) \left(1 - Q_{\tau}^k q^l
      \frac{Q_{\tau} q \tilde{t}}{t}
      \frac{x_i}{x_j}\right)}
\end{equation}
This measure is symmetric with respect to the exchange of $q$ and
$Q_{\tau}$, however the symmetry between $t$ and $\tilde{t}$ is lost.

\subsubsection{Torus $\widehat{A}_M$ measure}
\label{sec:virasoro-2}

We now have $M$ four-point amplitudes with edges totally
contracted.

\begin{figure}[h]
  \centering
  \includegraphics[width=5cm]{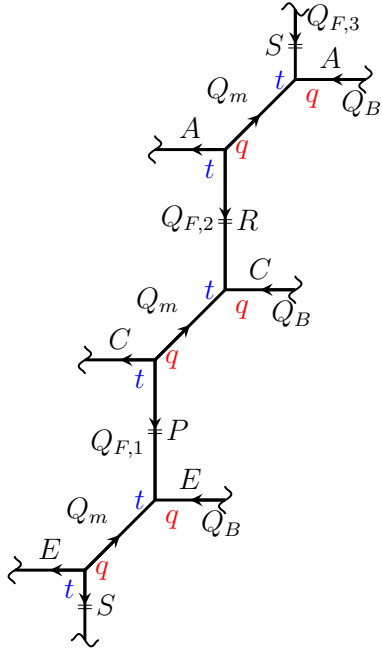}
  \caption{Three four-point amplitude with both vertical and
    horizontal ends identified to obtain the
    $\widehat{A}_2$ elliptic $q$-Selberg measure. Notice that
    $Q_{m,i}$ are now all the same, because of the constraints imposed
  by the closed hexagons formed by the identified edges.}
  \label{fig:Anhathat}
\end{figure}

The measure is a straightforward generalization of the
$\widehat{\mathfrak{u}}_1$ measure:
\begin{equation}
  \label{eq:35}
  \Delta^{(q,t,\tilde{t}, Q_{\tau})}_{\widehat{A}_M} (\vec{x}) =
  \frac{\prod_{a=1}^M \Delta^{(q,t,Q_{\tau})}(x_{(a)})}{\Delta^{(q,t,Q_{\tau})}(x_{(M)}
    , \tilde{t} x_{(1)}) \prod_{a=1}^{M-1}\Delta^{(q,t,Q_{\tau})}(x_{(a)}
    , \tilde{t} x_{(a+1)})}
\end{equation}
where
\begin{equation}
  \label{eq:36}
  \Delta^{(q,t,Q_{\tau})}(x) = \lim_{N \to \infty} \prod_{i \neq j} \prod_{k \geq 0}
  \prod_{l \geq 0} \frac{\left(1 - Q_{\tau}^k q^l
      \frac{x_i}{x_j}\right) \left(1 - Q_{\tau}^k q^l \frac{Q_{\tau} q}{t}
      \frac{x_i}{x_j}\right)}{\left(1 - Q_{\tau}^k q^l t
      \frac{x_i}{x_j}\right) \left(1 - Q_{\tau}^k q^l Q_{\tau} q
      \frac{x_i}{x_j}\right)}
\end{equation}
and
\begin{equation}
  \label{eq:37}
    \Delta^{(q,t,Q_{\tau})}(x,y) = \lim_{N \to \infty}
    \prod_{i=1}^{N_1} \prod_{j=1}^{N_2} \prod_{k \geq 0}
  \prod_{l \geq 0} \frac{\left(1 - Q_{\tau}^k q^l
      \frac{x_i}{y_j}\right) \left(1 - Q_{\tau}^k q^l \frac{Q_{\tau} q}{t}
      \frac{y_j}{x_i}\right)}{\left(1 - Q_{\tau}^k q^l t
      \frac{x_i}{y_j}\right) \left(1 - Q_{\tau}^k q^l Q_{\tau} q
      \frac{y_j}{x_i}\right)}
\end{equation}

\section{6d theory, affine Selberg integral and index vertex}
\label{sec:u1-theory-with}
In this section we investigate in more detail the affine Selberg
integrals first described in sec.~\ref{sec:ellipt-spher-block} and
point out an interesting natural generalization of them.

\begin{figure}[h]
  \centering
  \includegraphics[width=3cm]{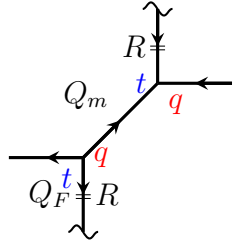}
  \caption{Compactified resolved conifold corresponding to free
    hypermultiplet in 6d or (in the spectral dual frame) to 5d $U(1)$
    theory with adjoint hypermultiplet.}
  \label{fig:101}
\end{figure}

We start with the simplest example and evaluate the partition function
of the trivial 6d theory with ``zero'' gauge group and only one free
multiplet using refined topological string. The corresponding toric
diagram is compactified along the preferred direction, as shown on
Fig.~\ref{fig:101}.

It is straightforward to obtain the (spectral dual) partition
function, which is given by the Nekrasov formula, i.e.\ the sum over
Young diagrams:
\begin{equation}
  \label{eq:103}
  Z^{\text{5d, adj}}_{U(1)} = \sum_R \left(\sqrt{\frac{t}{q}} Q_F
  \right)^{|R|} \frac{G_{RR}^{(q,t)}\left( \sqrt{\frac{q}{t}} Q_m
    \right)}{G_{RR}^{(q,t)}( 1 )}.
\end{equation}
Using the standard identities we can rewrite this sum over diagrams as
a contour integral of $q$-Selberg type:
\begin{equation}
  \label{eq:105}
  Z^{\text{5d, adj}}_{U(1)} = \lim_{Q \to 1} \lim_{N \to \infty} \mathcal{N}^{-1}  \oint_{\gamma}
  d^Nx \prod_{i=1}^N x_i^u \prod_{k \geq 0}\left[ \prod_{i \neq j} \frac{\left(1 - q^k \frac{x_i}{x_j}\right) \left(1 - t \tilde{t} q^k
        \frac{x_i}{x_j}\right)}{\left(1 - t q^k
        \frac{x_i}{x_j}\right)\left(1 - \tilde{t} q^k
        \frac{x_i}{x_j}\right)} \prod_{i=1}^N \frac{1 - \frac{Q q^k z_1}{x_i} }{1 - \frac{q^kz_1}{x_i}}\right]
\end{equation}
where $\tilde{t} = \sqrt{\frac{q}{t}} Q_m$, $q^u = \sqrt{\frac{t}{q}}
Q_F$ and $\mathcal{N}$ is the normalization constant. The poles are
enumerated by Young diagrams $R$ and are located at points
\begin{equation}
  \label{eq:106}
  x_{R,i} = z_1 q^{R_i} t^{N-i}.
\end{equation}
The integral~(\ref{eq:105}) differs from the ordinary $q$-Selberg
integral from sec.~\ref{sec:virasoro} in several respects:
\begin{enumerate}
\item The measure is of \emph{affine} Selberg type, more precisely of
  type $\widehat{\mathfrak{u}}_1$. This is in close analogy with the
  $A_n$ Selberg measure (see sec.~\ref{sec:w_n}), though the factor
  corresponding to the imaginary root contains extra $\tilde{t}$.

\item The integration measure in Eq.~(\ref{eq:105}) is explicitly
  symmetric between $t$ and $\tilde{t}$. This symmetry is completely
  unexpected from the point of view of the topological strings:
  $\tilde{t}$ in this framework represents the K\"ahler parameter of
  the resolved conifold while $t$ is the refinement parameter. Neither
  is this symmetry obvious from the corresponding 5d gauge theory:
  here $\tilde{t}$ is the mass of the adjoint multiplet and $t$ is one
  of the equivariant parameters.

\item The curious feature of the integral~(\ref{eq:105}) is the
  appearance of the special contour $\gamma$, which encircles only
  pole of the form $x_i = z_1 q^k t^l$ with $k,l \geq 0$ and excludes
  the poles $x_i = z_1 q^k t^l \tilde{t}^m$ with nonzero $m$. This
  choice of contour explicitly breaks the symmetry between $t$ and
  $\tilde{t}$ in the measure, so that the whole partition function is
  no longer symmetric. This can also be seen from the explicit
  infinite product formula for the partition function~(\ref{eq:103}),
  which has no symmetry between $\sqrt{\frac{q}{t}} Q_m$ and $t$.

\item The adjoint Nekrasov factors~(\ref{eq:103}) are non-vanishing for
  Young diagrams of arbitrary width. Thus, the number of integrations
  in~(\ref{eq:105}) is also infinite.

\item We insert an additional vertex operator $V_Q(z_1) =
  \prod_{i=1}^N \frac{1 - \frac{Q q^k z_1}{x_i} }{1 -
    \frac{q^kz_1}{x_i}}$ at $z_1$ to produce the poles of the
  necessary form~(\ref{eq:106}). We then take the limit $Q \to 1$ so
  that the extra factors cancel. One can naively think that the
  integral vanishes in this limit, since the poles in the denominator
  cancel with the zeroes of the numerator. However, we are in fact
  interested in the \emph{ratio} of the residues at the
  points~(\ref{eq:106}). More concretely, one sets $\mathcal{N} =
  \Res_{x_{\varnothing,i}} \mu(x)$, where $\mu(x)$ is the integrand,
  so that the sum over residues starts from the identity and is finite
  for $Q \to 1$. The value of the integral in this limit of course
  does not depend on the position $z_1$ of the vertex operator. We,
  therefore, view the additional factors in~(\ref{eq:105}) as a
  regularization.
\end{enumerate}

In the next section we modify the integral~(\ref{eq:105}) to include all
the poles and find that this is exactly
the equivariant instanton partition function of the 6d $(1,1)$ gauge
theory on $\mathbb{R}^6$.

\subsection{Extending the contour}
\label{sec:extending-contour}
Let us consider the following integral \cite{MMZ}
\begin{equation}
  \label{eq:101}
  \boxed{  Z_{\mathrm{ext}}^{U(1)} = \lim_{Q\to 1} \lim_{N\to \infty} \mathcal{N}^{-1}
    \oint\limits_{|x_i|=|z_1|+\epsilon} d^Nx \prod_{i=1}^N x_i^u \prod_{k \geq 0}\left[ \prod_{i \neq j} \frac{\left(1 - q^k
          \frac{x_i}{x_j}\right) \left(1 - t \tilde{t} q^k
          \frac{x_i}{x_j}\right)}{\left(1 - t q^k
          \frac{x_i}{x_j}\right)\left(1 - \tilde{t} q^k
          \frac{x_i}{x_j}\right)} \prod_{i=1}^N \frac{1 - \frac{Q q^k
          z_1}{x_i} }{1 - \frac{q^k z_1}{x_i}}
    \right]}
\end{equation}
where we assume $|q|<1$, $|t| > 1$, $|\tilde{t}| > 1$ and $\mathcal{N}
= \Res_{x_{\varnothing,i}} \mu(x)$. Notice the main difference
with~(\ref{eq:105}) --- the contour now encircles \emph{all} the poles
at $x_i = z_1 q^k t^{-l} \tilde{t}^{-m}$ with $k$, $l$, $m$
nonnegative.

\paragraph{Enumerating the poles.} Consider first the terms with $k=0$
in the integrand of Eq.~(\ref{eq:101}). Then for finite $N$ the
situation is completely analogous to the LMNS integrals for $U(1)$
gauge theory~\cite{Nekrasov:2002qd} (we will elaborate on this analogy
in sec.~\ref{sec:lmns-integrals-from}). One can see that one of the
variables $x_i$ should pick up a pole coming from the vertex
operator. The whole integrand is symmetric in $x_i$, so we can think
that this is the first variable, i.e.\ $x_1 = z_1$. Suppose then, that
we have already performed the first $l<N$ integrals and have picked
the poles, at $x_i = z_1 t^{1-n} \tilde{t}^{1-m}$, where $(n,m) \in Y'
\subset Y$. Then the next integration produces the following poles:
\begin{equation}
  \label{eq:1015}
  x_l = z_1 t^{1 -n \mp 1} \tilde{t}^{1-m}, \quad x_l = z_1 t^{1-n}
  \tilde{t}^{1-m \mp 1},\quad
  (n,m) \in Y'
\end{equation}
There are also zeroes:
\begin{equation}
\label{eq:1016}
x_l = z_1 t^{1-n} \tilde{t}^{1-m} \text{ (double)}, \quad x_l = z_1 t^{-n}
\tilde{t}^{-m},\quad x_l = z_1 t^{2-n}
\tilde{t}^{2-m},\quad
(n,m) \in Y'
\end{equation}
Some poles are canceled by zeroes, and only small portion of them
survives. In particular, if a point lies inside of $Y'$ and not on the
boundary of $Y'$, then there are exactly 4 poles and 4 zeroes which
cancel them. So, there is no pole to pick strictly inside $Y'$. On the
boundary the situation is a bit more subtle, and there are corner
contributions, recursion relations, etc.\ --- for details
see~\cite{Nakamura:2014nha}. However, when the dust settles one gets
the simple recipe, i.e.\ that each successive integration adds a box
so that the whole set of poles remains a Young diagram. All the poles
thus organize themselves into a Young diagram $Y$ with $N$ boxes:
\begin{equation}
  \label{eq:1017}
  x_{i,j} = z_1 t^{1-i}
\tilde{t}^{1-j}, \quad (i,j)\in Y.
\end{equation}

The terms with $k > 0$ have a simple effect --- each pole $x_{i,j}$ is
now shifted by $q^k$ with respect to the poles at $x_{i+1,j}$ or
$x_{i,j+1}$. Thus, the poles are now labelled by a \emph{plane
  partition} (3d Young diagram) $\pi$ with \emph{floor area} $N$, so
that $x_{i,j} = z_1 q^{\pi_{i,j}-1} t^{1-i} \tilde{t}^{1-j}$.

For infinite $N$ a slightly different picture is more convenient. Let
us demolish the ``stylobate'' of $\pi$ -- i.e.\ reduce the height of
each column by one. The resulting plane partition has floor area
\emph{less or equal} to $N$ and will be denoted by $\tilde{\pi}$, so
that $\tilde{\pi}_{i,j} = \pi_{i,j}-1$.

\paragraph{Computing the residues.} Let us now compute the residues at
the poles. More concretely, since the normalization constant is the
contribution of the pole corresponding to the empty diagram, we are
actually computing the \emph{ratio} of the residues (the same trick
was used in~\cite{Aganagic:2013tta}):
\begin{equation}
  \label{eq:107}
  Z_{\mathrm{ext}}^{U(1)} = \sum_{\tilde{\pi}} \frac{\Res_{x_{\tilde{\pi}}}
    \mu(x)}{\Res_{x_{\varnothing}} \mu(x)} = \sum_{\pi}
  \frac{\mu(x_{\pi})}{\mu(x_{\varnothing})}
\end{equation}
We find that the residues have the following plethystic form:
\begin{equation}
  \label{eq:108}
  \frac{\mu(x_{\pi})}{\mu(x_{\varnothing})} = q^{u|\tilde{\pi}|} \exp \left\{ -\sum_{n
      \geq 1} \frac{1}{n} \left[ \frac{(1 - t^n)(1-\tilde{t}^n)}{1-q^n}
      \left( p_n(x_{\tilde{\pi}}) p_{-n}(x_{\tilde{\pi}}) - p_n(x_{\varnothing})
        p_{-n}(x_{\varnothing})\right) 
  \right] \right\}.
\end{equation}
where $p_n(x)$ are power sum symmetric functions. We can take the
limit $N \to \infty$ in each term, and the power sums become
\begin{multline}
  \label{eq:109}
  p_n(x_{\tilde{\pi}}) = \sum_{i,j \geq 1}^{\infty} q^{n
    \tilde{\pi}_{i,j}} t^{n(1-i)} \tilde{t}^{n(1-j)} = \frac{1}{(1 -
    t^{-n})(1 - \tilde{t}^{-n})} + (q^n-1) \sum_{(i,j,k) \in \pi}
  q^{n(k-1)} t^{n(1-i)}
  \tilde{t}^{n(1-j)}=\\
  = \frac{1}{(1-t^{-n}) (1-\tilde{t}^{-n})} \left[ 1 -
    (1-q^n)(1-t^{-n})(1-\tilde{t}^{-n}) \ch_{\tilde{\pi}} (t^{-n},
    \tilde{t}^{-n} ,q^n) \right],
\end{multline}
where the $\ch_{\tilde{\pi}} (q_1, q_2, q_3) = \sum_{(i,j,k)\in
  \tilde{\pi}} q_1^{i-1} q_2^{j-1} q_3^{k-1}$. Finally, the value of
the residues is given by
\begin{equation}
  \label{eq:1010}
  \frac{\mu(x_{\tilde{\pi}})}{\mu(x_{\varnothing})} = q^{u|\tilde{\pi}|} \exp \left\{ -\sum_{n
      \geq 1} \frac{1}{n} \frac{E_{\tilde{\pi}}(t^{-n}, \tilde{t}^{-n}, q^n) E_{\tilde{\pi}}(t^n,
      \tilde{t}^n, q^{-n}) - 1}{(1 -
      t^{-n})(1-\tilde{t}^{-n})(1-q^n)} \right\},
\end{equation}
where $E_{\tilde{\pi}}(q_1, q_2, q_3) = 1 - (1-q_1)(1-q_2)(1-q_3)
\ch_{\tilde{\pi}} (q_1, q_2, q_3)$. In s.~\ref{sec:u1-theory-with} we
have found an unexpected symmetry between $t$ and $\tilde{t}$ in the
integration measure. However, the choice of the special contour
$\gamma$ did not respect this symmetry. Choosing the contour
$|x|=|z_1|+\epsilon$ we not only restore the symmetry $t
\leftrightarrow \tilde{t}$, but also get a free bonus. The partition
function~(\ref{eq:1010}) is actually completely symmetric in $t^{-1}$,
$\tilde{t}^{-1}$ and $q$. This is part of our motivation for extending
the integration contour. Topological string theory does not give a
clue about the origin of this extra symmetry. In the next section we
will argue that it can be understood from the six-dimensional point of
view. More concretely, we will show that Eq.~(\ref{eq:1010}) exactly
reproduces the sum over fixed points in the instanton moduli space of
the 6d $\mathcal{N}=(1,1)$ $U(1)$ gauge theory.

\subsection{6d $\mathcal{N}=(1,1)$ $U(1)$ theory and the index vertex}
\label{sec:u1-theory-six}

The $\mathcal{N}=(1,1)$ gauge theory has the maximal possible
supersymmetry in six dimensions. It can be straight-forwardly obtained
from dimensional reduction of 10d $\mathcal{N}=1$ gauge theory. In
flat Euclidean space the supersymmetry can be equivariantly twisted
using the maximal torus of isometries of $\mathbb{R}^6$, i.e.\
$U(1)^3$. The equivariant partition function, therefore, depends on
three equivariant parameters $q_{1,2,3} = e^{\epsilon_{1,2,3}}$. The
equivariant integrals over the $Q$-closed field configurations
localize on the fixed points, which are labelled by plane partitions,
so that the instanton partition function is given by the sum over
fixed points each taken with its equivariant index\footnote{In the
  case of $U(1)$ gauge theory there is also an explicit formula for
  the whole sum in terms of an infinite product. However, since we are
  also interested in the generalization to $U(N)$ (in which case the
  infinite product formula is lacking), we will not write it down
  here.}~\cite{Iqbal:2003ds},~\cite{Szabo:2015wua}
\begin{equation}
  \label{eq:104}
  Z^{\mathrm{6d} (1,1) U(1)}_{\mathrm{inst}} = \sum_{\pi} \Lambda^{|\pi|} \mathrm{Ind}_{\pi}(q_1,q_2,q_3),
\end{equation}
where $\Lambda$ is the coupling constant. The index of each fixed
point is given by the product of $U(1)^3$ weights:
\begin{equation}
  \label{eq:1011}
  \mathrm{Ind}_{\pi}(q_1, q_2, q_3) = \prod_i (1 - e^{w_i(q, \pi)})^{\pm 1}
\end{equation}
It is convenient to write the weights in the plethystic form:
\begin{equation}
  \label{eq:1012}
  \mathrm{Ind}_{\pi}(q_1, q_2, q_3) = \exp \left[ \sum_{n\geq 1}
    \frac{1}{n}  \ch_{T\pi} (q_1^n, q_2^n , q_3^n )\right]
\end{equation}
where the character of the tangent space to the fixed point is given
by
\begin{equation}
  \label{eq:1013}
  \ch_{T\pi} (q_1, q_2 , q_3) = \sum_i \mp e^{w_i(q, \pi)}
\end{equation}
These characters have been found in~\cite{Nekr-QUARKS} and are given
by
\begin{equation}
  \label{eq:1014}
  \ch_{T\pi} (q_1, q_2 , q_3) = \frac{1 - E_{\pi}(q_1, q_2, q_3)
    E_{\pi}(q_1^{-1}, q_2^{-1}, q_3^{-1})}{(1 - q_1^{-1})(1-q_2^{-1})(1-q_3^{-1})}.
\end{equation}

One immediately sees that this sum over fixed points is exactly the
same as the sum over poles in Eq.~(\ref{eq:1010}) provided one makes the
following identifications:
\begin{equation}
  \label{eq:102}
  \boxed{
    \begin{array}{rcl}
      q_1 &=& t^{-1},\\
      q_2 &=& \tilde{t}^{-1} = \sqrt{\frac{q}{t}} Q_m,\\
      q_3 &=& q,\\
      \Lambda &=& q^u = \sqrt{\frac{t}{q}}Q_F.
    \end{array}
}
\end{equation}
At this point several remarks are in order.
\begin{enumerate}
\item Partition function, similar to~(\ref{eq:104}) was used
  in~\cite{Iqbal:2003ds} to find the partition function of topological
  strings on the $\mathbb{C}^3$ patch of a toric CY manifold. The main
  difference from~(\ref{eq:104}) was that the plane partitions were
  allowed to have infinite ``legs'' along the three coordinate axes,
  so that the resulting vertex depended on the three Young diagrams in
  the asymptotics.

\item Using the map~(\ref{eq:102}) one can understand the remarkable
  symmetry between $t$, $\tilde{t}$ and $q$ in the enlarged
  integral as the symmetry between three equivariant parameters in
  $\mathbb{R}^6$, and eventually relate it to the action of the Weyl
  group of $SO(6)$.
\end{enumerate}

\subsection{Generalizing to $U(N)$/quiver of $U(1)$ groups}
\label{sec:generalizing-un}
\begin{figure}[h]
  \centering
  \includegraphics[width=9cm]{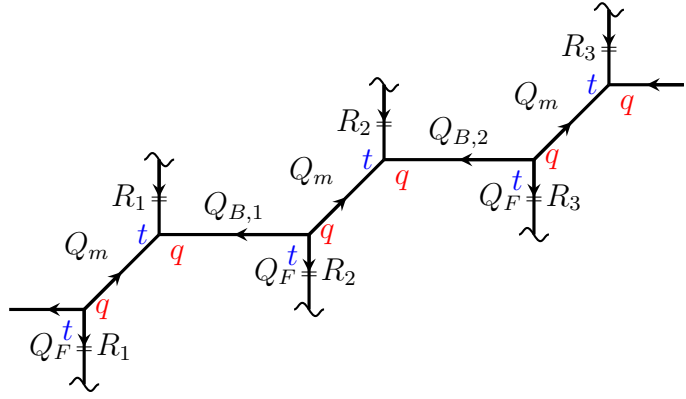}
  \caption{Stack of compactified resolved conifolds corresponding to
    5d $U(3)$ theory with adjoint hypermultiplet.}
  \label{fig:102}
\end{figure}

Generalization to linear quiver of $U(1)$ groups in 6d or equivalently
to 5d $U(N)$ gauge theory is straightforward. We consider a stack of
$N$ compactified resolved conifolds as shown on Fig.~\ref{fig:102}.

The partition function is just 5d $U(N)$ Nekrasov function with
adjoint multiplet. It has an integral representation quite similar
to~\eqref{eq:105}, the only difference is that there are $N$ vertex
operator insertions at points $z_1$,\dots, $z_N$:
\begin{equation}
  \label{eq:23}
  Z^{\text{5d, adj}}_{U(N)} = \lim_{Q_a \to 1} \lim_{M \to \infty} \mathcal{M}^{-1}  \oint_{\gamma}
  d^Mx \prod_{i=1}^M x_i^u \prod_{k \geq 0}\left[ \prod_{i \neq j} \frac{\left(1 - q^k \frac{x_i}{x_j}\right) \left(1 - t \tilde{t} q^k
        \frac{x_i}{x_j}\right)}{\left(1 - t q^k
        \frac{x_i}{x_j}\right)\left(1 - \tilde{t} q^k
        \frac{x_i}{x_j}\right)} \prod_{a=1}^N\prod_{i=1}^M \frac{1 - \frac{Q_a q^k z_1}{x_i} }{1 - \frac{q^kz_a}{x_i}}\right]
\end{equation}

The poles of the integral are now enumerated not by one partition, but
by an $N$-tuple of partitions $\vec{R} = (R_1, \ldots, R_N)$, so that
$x_{\vec{R}} = \tilde{z}_a q^{R_{a,i}} t^{1-i}$, where $\tilde{z}_a =
z_a t^{N_a-1}$.

Just as in the $U(1)$ case, the contour $\gamma$ can be extended to
encircle the additional poles, so that the whole partition function is
the sum over $N$-tuples of plane partitions, $\vec{\pi} = (\pi_1,
\ldots, \pi_N)$, and $x_{\vec{\pi}} = \tilde{z}_a q^{\pi_{a,i,j}} t^{1-i}
\tilde{t}^{1-j}$. Moreover, the sum over residues exactly reproduces
the localization formula for the 6d $\mathcal{N}=(1,1)$ $U(N)$
theory. For the (scaled) sum over residues we have:
\begin{multline}
  \label{eq:38}
  Z_{\mathrm{ext}}^{U(N)} = \sum_{\vec{\pi}}
  \frac{\Res_{x_{\vec{\pi}}} \mu(x)}{\Res_{x_{\varnothing}} \mu(x)} =
  \sum_{\vec{\pi}}
  \frac{\mu(x_{\vec{\pi}})}{\mu(x_{\varnothing})} =\\
  =\sum_{\vec{\pi}} q^{u |\vec{\pi}|} \exp \left\{ -\sum_{n \geq 1}
    \frac{1}{n} \left[ \frac{(1 - t^n)(1-\tilde{t}^n)}{1-q^n} \left(
        p_n(x_{\vec{\pi}}) p_{-n}(x_{\vec{\pi}}) -
        p_n(x_{\varnothing}) p_{-n}(x_{\varnothing})\right)\right]
  \right\}=\\
  = \sum_{\vec{\pi}} q^{u |\vec{\pi}|} \exp \left\{ -\sum_{n \geq 1}
    \frac{1}{n} \left[ \frac{(1 - t^n)(1 - \tilde{t}^n)}{1 - q^n}
      \sum_{a,b=1}^N z_a^n z_b^{-n} \left( p_n(x_{\pi_a})
        p_{-n}(x_{\pi_b}) - p_n(x_{\varnothing})
        p_{-n}(x_{\varnothing})\right)\right] \right\}
\end{multline}
Whereas the localization in the gauge theory looks similar
to~\eqref{eq:104}:
\begin{equation}
  \label{eq:39}
  Z_{\mathrm{inst}}^{6d (1,1) U(N)} = \sum_{\vec{\pi}}
  \Lambda^{|\vec{\pi}|} \mathrm{Ind}_{\vec{\pi}} (q_1, q_2, q_3),
\end{equation}
except the index now depends on $N$-tuple of plane partitions and $N$
Coulomb moduli of the theory:
\begin{equation}
  \label{eq:40}
  \mathrm{Ind}_{\vec{\pi}} (q_1, q_2, q_3) = \exp \left\{ -
    \sum_{n\geq 1} \frac{1}{n} \frac{W(\vec{a}^n,\vec{q}^n) W(\vec{a}^{-n},\vec{q}^{-n}) - E_{\vec{\pi}}(\vec{a}^n,\vec{q}^n) E_{\vec{\pi}}(\vec{a}^{-n},\vec{q}^{-n})}{(1 - q_1^{-n})(1 - q_2^{-n})(1 - q_3^{-n})} \right\},
\end{equation}
where
\begin{equation}
  W(\vec{a}, \vec{q}) = \sum_{p=1}^N a_p, \qquad E(\vec{a}, \vec{q}) =
  \sum_{p=1}^N \sum_{(i,j,k) \in \pi_a} a_p q_1^{i-1} q_2^{j-1} q_3^{k-1}
\end{equation}
Notice, that for $N=1$ the dependence of the index on $a_p$ drops out,
as it should (there are no Coulomb moduli in the abelian
theory). Partition functions~\eqref{eq:38} and~\eqref{eq:39}
manifestly coincide if we make the following identification of the
gauge theory and topological string parameters:
\begin{equation}
  \boxed{
    \begin{array}{rcl}
      q_1 &=& t^{-1}\\
      q_2 &=& \tilde{t}^{-1} = \sqrt{\frac{q}{t}} Q_m\\
      q_3 &=& q\\
      \Lambda &=& q^u = \sqrt{\frac{t}{q}} Q_F\\
      a_p &=& \tilde{z}_p
    \end{array}
}
\end{equation}
Notice that the extended integral in the $U(N)$ case is still
symmetric in $q$, $t^{-1}$ and $\tilde{t}^{-1}$.

\subsection{AGT: LMNS integral from the extended DF integral}
\label{sec:lmns-integrals-from}

The LMNS integrals and DF integrals describe respectively the instanton partition functions and conformal blocks. Their equivalence is known as AGT duality and it is usually seen as a non-trivial  integral transform of the Habbard-Stratonovich type \cite{MMSha}. In \cite{MMZ}, we suggested that it can actually be raised to an explicit {\it symmetry}. Namely, our extended ``six-dimensional'' integral is a certain generalization of both integrals, which can be turned into either the $q$-deformed version of the
DF integral or the 5d LMNS integral in suitable limits. These two limits are given by $\tilde{t} \to 0$ and $q\to 0$ respectively, and the resulting expressions are related by transformations $t^{-1}\leftrightarrow \tilde{t}^{-1}
\leftrightarrow q$. This is an exact and explicit symmetry of the integral. This limiting procedure straightforwardly
generalizes to $U(N)$ theory. Since the initial extended integral is
still symmetric in the three equivariant parameters, DF and LMNS
representations are exactly equivalent.

One can go further and study the 4d limit of the gauge theory. In this
case $q$-deformed DF integral turns into the ordinary DF or
beta-ensemble integral, and 5d LMNS integral reduces to the ordinary
LMNS one. However, at this level the symmetry is no longer explicit and looks almost like a miracle, if one does not know that the integrals came from an explicitly symmetric 6d expression.

\section{Spectral dualities and elliptic integrable systems}

\subsection{Generalities\label{gen}}

According to \cite{GKMMM}, the hierarchy of physical theories
associated with brane configurations in string and low-energy
Yang-Mills models actually begins with much simpler integrable systems,
and Seiberg-Witten theory is exactly the one, which captures the information
available at this level of description.
Moreover, the Nekrasov deformation of Seiberg-Witten theory corresponds to a quantization
of these systems \cite{NSM}, or, more precisely, a lifting from quasiclassical to
full-fledged $\tau$-function theory, what, within the integrable theory context,
is a general straightforward procedure, which does not require any additional
information.

This circle of ideas was further developed and exploited in numerous works.
It is now well known that, within this context, the
5d gauge theory with adjoint hypermultiplet corresponds to
the elliptic Ruijsenaars system \cite{Braden}, the 6d linear quiver theory gives the XYZ spin chain \cite{MirMar}
and the 6d gauge theory with
adjoint matter is described by the double elliptic integrable system \cite{Dell}.
In other words, for generic toric diagrams, which we consider in this paper,
we have the following table:

\begin{center}
  \begin{tabular}{c|c|c|c}
    & Dimension & Sphere & Torus\\
    \hline
    algebra&4d&XXX chain &Elliptic Calogero\\
  $q$-deformed algebra & 5d &XXZ chain & Elliptic Ruijsenaars\\
  doubly-deformed algebra & 6d& XYZ chain& Double elliptic\\
\end{tabular}
\end{center}

\noindent
All these systems should be related by various dualities,
of which the most non-trivial are spectral dualities, interchanging
vertical and horizontal directions in toric diagrams.
They make XYZ spin chain equivalent to the Ruijsenaars
system and different double elliptic systems are also equivalent.
In the integrable system context, this equivalence was first realized by K.Hasegawa \cite{HV} (in fact, in some part by E.Sklyanin \cite{Skl}). We describe it briefly in this section, postponing the details until a separate paper \cite{MMZZ}.

Let us start simply by counting the parameters of the spectral dual
gauge theories.

$SU(N)$ theory with fundamental matter in 6d has the following
parameters: Coulomb moduli $a_i$, $i=1,\ldots ,N-1$, masses of the
fundamental hypermultiplets $m_i^{\pm}$, $i=1,\ldots , N$, the
coupling constant $\Lambda$ and the two radii of the compactified
dimensions $R_5$ and $R_6$. There is also one feature unique to 6d
theories: the masses cannot all be set independently, but there is one
condition on them. This makes the total number of parameters $3N+1$.

$SU(2)^N$ necklace quiver theory in 5d has the following parameters
parameters: Coulomb moduli $\widetilde{a}_i$, $i=1,\ldots, N$, masses
of the bifundamentals $\widetilde{m}_i$, $i=1,\ldots, N$, coupling
constants $\widetilde{\Lambda}_i = 1,\ldots, N$, and the radius of the
fifth dimension $R_5$. In total one gets $3N + 1$ parameters.

This counting can also be seen on the toric diagram. We consider the
example of $SU(2)^4$:

\begin{center}
\includegraphics[width=0.8\textwidth]{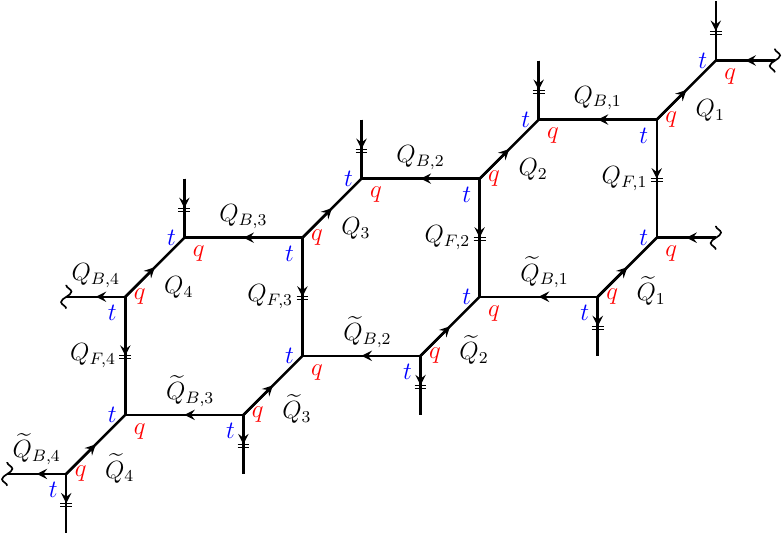}
\end{center}

The amplitude depends on $5\times 4 = 20$ K\"ahler parameters written
explicitly on the diagram and also on the radius $R_5$ of the M-theory
circle. Each hexagon on the diagram enforces a pair of constraints on
the K\"ahler parameters of its edges:
\begin{gather}
  Q_{F,i} \widetilde{Q}_i = Q_{F,i+1} Q_{i+1},\qquad i=1,\dots, 4,\\
  Q_{B,i} Q_{i+1} = \widetilde{Q}_i \widetilde{Q}_{B,i},\qquad
  i=1,\ldots 4.
\end{gather}
where we set $Q_{F,i+4} = Q_{F,i}$, etc. In total we get $5 \times 4 -
2\times 4 = 3 \times 4 = 12$ independent K\"ahler parameters and
$R_5$, which agrees with the gauge theory counting, which also gives
$3N+1 = 3 \times 4 + 1 = 13$.

\subsection{XYZ chain}
\label{sec:xyz-chain}

$SU(N)$ gauge theory with $2N$ fundamental
hypermultiplets\footnote{For other numbers of multiplets there is a
  gauge anomaly.} in 6d corresponds \cite{MirMar} to the Sklyanin $N$-site $U_{q,t}(\mathfrak{gl}_2)$ XYZ
spin chain \cite{Skl}. The transfer matrix of the spin chain is written in terms
of Lax matrices, residing on each site of the chain with the corresponding inhomogeneity $\xi_i$:
\begin{equation}
  T(\xi) = L^{(N)}(\xi-\xi_N)\ldots L^{(1)}(\xi-\xi_1),
\end{equation}
where
\be\label{39}
L(\xi) = S^0\cdot {\bf 1} + i\sum_{a=1}^3 W_a(\xi)S^a
\sigma_a
\ee
$\sigma_a$ are the Pauli matrices, and
\be\label{sklyatheta}
W_a(\xi) = \sqrt{e_a - \wp\left({\xi}|\tau\right)} =
i\frac{\theta'_{\ast}(0)\theta_{a+1}\left({\xi}\right)}{\theta_{a+1}(0)
\theta_*(\xi)}
\ee
where $\theta_a (x)$ is the standard Jacobi $\theta$-function and $e_i$ are values of the Weierstrass function $\wp (\xi)$ at the half-periods.
The dynamical variables $S^0, S^a$ form the (classical)
Sklyanin algebra \cite{Skl}:
\be\label{sklyal}
\left\{S^a, S^0\right\} = -i\left(e_b - e_c\right)S^bS^c\ \  \ \ \ \ \ 
\left\{S^a, S^b\right\} = -iS^0S^c
\ee
with the obvious notation: $abc$ is the triple $123$ or its cyclic
permutations.

For this $U_{q,t}(\mathfrak{gl}_2)$ chain there are two Casimir operators, i.e. one degree of freedom remaining per site, which means that there are totally $N$ action variables, which correspond to $N$ Coulomb moduli $a_i$. However, one can consider vanishing the full momentum of the system (which corresponds to removing the $U(1)$-factor in the gauge theory), in this way, we are left with $N-1$ Coulomb moduli. The Casimirs and inhomogeneities are combined into $2N$ parameters of the fundamental hypermultiplet masses \cite[(4.14)-(4.15)]{MirMar}. In fact, there is a restriction imposed on the sum of all masses \cite{MirMar}, thus, there are $2N-1$ parameters. This matches the counting of degrees of freedom above.

The 5d and 4d reductions of this theory is described by the XXZ and XXX chain respectively.

\subsection{Elliptic spin Ruijsenaars system}
\label{sec:ellipt-spin-ruijs}

The $SU(N)$ gauge theory with the adjoint hypermultiplet in 5d corresponds \cite{Braden} to the elliptic Ruijsenaars system \cite{Rui} given by the Lax operator
\be
L_{ij}(\xi) = c(\xi|\epsilon)e^{p_i}\prod_{k\ne i}\sqrt{\wp (q_{ik})-\wp (\epsilon)}{F(q_{ij}|\xi)\over F(q_{ij}|\epsilon)}
\ \ \ \ \ \ \ \ F(q|\xi)={\theta (q+\xi)\theta'(0)\over \theta(q)\theta(\xi)}
\ee
where $c(\xi|\epsilon)$ is a normalization factor which has to be chosen in a convenient way.

In order to extend this theory to the product of gauge groups, one first has to consider the spin elliptic Ruijsenaars system \cite{KZ}.
Then, the corresponding Lax operator is just 
\be
L_{ij}(\xi)\sim S_{ij}e^{p_i}\prod_{k\ne i}\sqrt{\wp (q_{ik})-\wp (\epsilon)}{F(q_{ij}|\xi)\over F(q_{ij}|\epsilon)}
\ee
with more dynamical variables: spins $S_{ij}$.

The next step is to extend it further to multi-point system. The Lax operator in this case becomes much more involved and so does the Poisson bracket of the spin variables $S^a_{ij}$ \cite{MMZZ}. For the sake of simplicity, we write down here only the 4d case, when the system is multi-point Calogero system, and the formulas are much more compact, while the 5d formulas can be found in \cite{MMZZ}. In the 4d case, the gauge theory has the gauge group $U(1)\times SU(N)^k$ and contains $k$ matter bifundamentals \cite{Witten:1997sc}. On the integrable side, the multi-point spin Calogero system is described by the Lax operator given on a torus with $k$ marked points $w_a$, \cite{mCal}:
\be\label{LG}
L_{ij}(\xi) = \delta_{ij}\left(p_i+\sum_a S^a_{ii}\zeta(\xi-w_a)\right) + (1-\delta_{ij}) \sum_aS_{ij}^aF(q_{ij}|\xi-w_a)
\ee
where the spin variables satisfy the Poisson bracket
\be
\{S_{ij}^a,S_{kl}^b\}=\delta_{ab}\left(S_{il}^a\delta_{jk}-S^a_{jk}\delta_{il}\right)
\ee
and there is an additional constraint
\be
\sum_a S^a_{ii}=0
\ee
The Poisson bracket is non-degenerate upon reducing the spin matrices to the orbits of $\mathfrak{gl}_N$.
Thus, the system is characterized by the three integers: the number of particles $N$, the number of marked points $k$ and the parameter of the orbit $l$.

\subsection{Spectral duality}

The spectral duality, which we mentioned in ss.\ref{1.1} and \ref{gen}, connects the Seiberg-Witten theories in 6d and 5d gauge theories. At the level of integrable systems, it was established by K.Hasegawa \cite{HV} (see also a trigonometric version of the correspondence in \cite{AHZ}) and claims an equivalence of the elliptic multi-point spin
Ruijsenaars system given by $(N,k,l)$ and the elliptic spin chain on $k$ sites, given by the $l$-orbit of the Sklyanin-
Odesskii-Feigin $\mathfrak{gl}_N$ \cite{OF}. In the particular case of $l = 1$ (the orbit of minimal dimension), one obtains the
duality between $SU(N)^k$ theory with fundamental matter in 6d and $SU(k + 1)^N$ necklace quiver theory in 5d.
Since this is a subject of its own value, we discuss implications of this Hasegawa correspondence between integrable systems in some more detail in a separate paper \cite{MMZZ}.

This duality can be lifted to the 6d theories with adjoint matter, which are described by the double elliptic
integrable systems \cite{Dell}. These systems have not been studied in full yet, because of a very involved structure (see \cite{amitoappear} for some new advances).  As usual, they appear from
explicit expressions for  partition function in section 3 in quasiclassical limit $\epsilon_1\sim\epsilon_2\longrightarrow 0$, while in Nekrasov-Shatashvili limit \cite{NSM}  (when only $\epsilon_2\longrightarrow \infty$)  we get their straightforward quantization, when the spectral curve is substituted by a Baxter equation (quantum spectral curve). Analysis of these limits could help to describe the full integrable double elliptic system.

\section{Conclusion}

In this paper, we attempted to describe the Seiberg-Witten/Nekrasov theory for
the most general model associated with an arbitrary $(p,q)$-web toric diagram
(the tropical limit of the spectral curve).
In the gauge theory language, this corresponds to 6d theory, in the integrable system language
to the double elliptic system.
We explained that the recent advances in the theories of Dotsenko-Fateev integrals and
topological integrals provide a straightforward dictionary for conversion
between the pictorial language of toric diagrams (spectral curves) and
the Young-diagram expansions for the
Nekrasov functions, and this dictionary gets remarkably simple at this
most general level.
Numerous string dualities have non-trivial realizations in all the languages,
and they turn into precise equivalences between the Nekrasov functions,
reflecting precise equivalences between the integrable systems.
The most interesting of the latter are spectral dualities between the integrable systems
of spin chain (XYZ) and Calogero-Ruijsenaars types.

An enormous amount of work is still necessary to polish this description.
Most important is to find an adequate extension of the matrix model formalism, which would make
the dualities transparent.
In its usual form, from \cite{FDagt} to \cite{Nek,Kimura:2015rgi},
it treats differently the horizontal and vertical directions in toric diagrams.
At the same time, it is the only approach which straightforwardly provides
the entire set of Ward identities (Virasoro/$W_N$-constraints, loop equations) for the Nekrasov expansions
and the AGT related conformal blocks.
Desired is an efficient formalism, where explicit are both the perturbative Ward identities
and non-perturbative dualities.
We hope that this paper clearly demonstrates that such a description is fully consistent,
but an adequate formalism still needs to be found.

\section*{Acknowledgements}

We are grateful to A.Zotov for the discussions of multipoint elliptic integrable systems. 

Our work
is partly supported by grants 15-31-20832-Mol-a-ved (A.Mor.), 15-31-20484-Mol-a-ved (Y.Z.), by RFBR grants 16-01-00291 (A.Mir.)
and 16-02-01021 (A.Mor. and Y.Z.), by joint grants 15-51-50034-YaF,
15-51-52031-NSC-a, 16-51-53034-GFEN, by the Brazilian National
Counsel of Scientific and Technological Development (A.Mor.).

\appendix

\section{Five-dimensional Nekrasov functions and AGT relations}
\label{sec:five-dimens-nekr}
The Nekrasov partition function for the $U(N)$ theory with $N_f = 2N$
fundamental hypermultiplets is given by
\begin{equation}
\label{eq:43}
Z_{\mathrm{Nek}}^{5d,\, U(N)} = \sum_{\vec{A}} \Lambda^{|\vec{A}|}
\frac{z_{\mathrm{fund}} (\vec{A}, \vec{m}^{+}, \vec{a})
  z_{\overline{\mathrm{fund}}} (\vec{A}, \vec{m}^{-}, \vec{a})
}{z_{\mathrm{vect}}(\vec{A},\vec{a})} = \sum_{\vec{A}} \Lambda^{|\vec{A}|}
\frac{\prod_{i=1}^{N}\prod_{f=1}^{N} f_{A_i}^{+} (q^{m_f^{+} + a_i}) f_{A_i}^{-} (q^{m_f^{-} + a_i})
}{z_{\mathrm{vect}}(\vec{A},\vec{a})}\,,
\end{equation}
where $f_A^{\pm} (q^x) = \prod_{(i,j) \in A} \left(1 - q^{\pm x}
  t^{\pm ( i - 1)} q^{\mp (j - 1 )} \right)$,
$z_{\mathrm{vect}}(\vec{A},\vec{a}) = \prod_{i,j=1}^{N} G^{(q,t)}_{A_i
  A_j}(q^{a_i - a_j})$ and
\begin{multline}
\label{eq:42}
  G^{(q,t)}_{AB} (q^x)= \prod_{(i, j) \in A} \left( 1 - q^x q^{A_i - j}
    t^{B^{\mathrm{T}}_j - i + 1} \right) \prod_{(i,j) \in B}\left(1 -
    q^x q^{-B_i + j - 1} t^{-A^{\mathrm{T}}_j + i} \right) =\\
 = \prod_{(i, j) \in B} \left( 1 - q^x q^{A_i - j}
    t^{B^{\mathrm{T}}_j - i + 1} \right) \prod_{(i,j) \in A}\left(1 -
    q^x q^{-B_i + j - 1} t^{-A^{\mathrm{T}}_j + i} \right),
\end{multline}
in particular
\begin{gather}
  \label{eq:76}
  G^{(q,t)}_{A \varnothing} (q^x) = \prod_{(i,j)\in A} \left( 1 - q^x
    q^{j-1} t^{1-i} \right) = f^{-}_A(q^{-x}),\\
  G^{(q,t)}_{\varnothing A} (q^x) = \prod_{(i,j)\in A} \left( 1 - q^x
    q^{1-j} t^{i-1} \right) = f^{+}_A(q^x),
\end{gather}

We will write $a$ instead of $\vec{a}=(a,-a)$ for $N=2$. The AGT
relations written in terms of the DF or Selberg integral parameters
for $N=2$ are:
\begin{gather}
  u_{+} = m_1^{+} - m_2^{+} - 1 + \beta\,, \qquad \qquad u_{-} = -1 +
  \beta -2a \,, \notag\\
  v_{+} = - m^{+}_1 - m^{+}_2 \,, \qquad \qquad v_{-} =
  - m_1^{-} - m_2^{-}\,, \label{eq:41}\\
  \beta n_{+} = -a + m^{+}_2\,, \qquad \qquad \beta n_{-} = a +
  m^{-}_2\,,\notag
\end{gather}
where $a_1 = - a_2 = a$. Masses $m_a$, vevs $a_i$, radius $R_5$ of the
fifth dimension and $\epsilon_{1,2}$ all have dimensions of mass. In
this paper we set the overall mass scale so that $\epsilon_1 = -b^2$,
$\epsilon_2 = 1$ and $q = e^{-R_5}$. The $t$ parameter in Macdonald
polynomials is related to $q$ by $t = q^{\beta}$ with $\beta = b^2$.

More generally, one can consider quiver gauge theories with gauge
groups $U(N)^k$ and bifundamental matter hypermultiplets. The
corresponding Nekrasov function is
\begin{multline}
  \label{eq:61}
  Z_{\mathrm{Nek}}^{5d,\, U(N)^k} = \sum_{\vec{Y}_a} \Lambda_1^{|\vec{Y}_1|} \cdots
  \Lambda_k^{|\vec{Y}_k|} \prod_{f=1}^N \prod_{i=1}^N
  f^{+}_{Y_{1,i}}\left(q^{m_f^{+} + a_{1,i}} \right)
  \frac{1}{z_{\mathrm{vec}}(\vec{Y}_1, \vec{a}_1)}
  z_{\mathrm{bifund}}\left(\vec{Y}_1, \vec{Y}_2, \vec{a}_1, \vec{a}_2,
  m_{\mathrm{bifund},1}\right) \cdots\\
  \cdots z_{\mathrm{bifund}}\left(\vec{Y}_{k-1}, \vec{Y}_k, \vec{a}_{k-1},
  \vec{a}_k, m_{\mathrm{bifund},k-1}\right)
  \frac{1}{z_{\mathrm{vec}}(\vec{Y}_k , \vec{a}_k)} \prod_{f=1}^N
  \prod_{i=1}^N f^{-}_{Y_{k,i}}\left(q^{m_f^{-} + a_{k,i}} \right)
\end{multline}
where the bifundamental contribution is given by
$z_{\mathrm{bifund}}(\vec{Y}, \vec{W}, \vec{a}, \vec{b}, m) =
\prod_{i=1}^N \prod_{j=1}^N G_{Y_i W_j}^{(q,t)} \left( q^{a_i - b_j -
    m} \right)$.

\section{Useful formulas}
\label{sec:useful-formulas}
\begin{gather}
  \label{eq:2}
  \prod_{i,j\geq 1} \frac{1 -  Q q^{j - W_i - \frac12} t^{i -
      Y^{\mathrm{T}}_j - \frac12}}{1 - Q q^{j-\frac12} t^{i-\frac12}}
  = \prod_{i,j,k\geq 1} \frac{\left( 1 - \sqrt{\frac{q}{t}} Q q^{k + Y_i - W_j }
      t^{j - i
       + 1 } \right) \left(1 - \sqrt{\frac{q}{t}} Q
    q^k t^{j - i}\right)}{\left(1 - \sqrt{\frac{q}{t}} Q
    q^{k + Y_i - W_j} t^{j - i}\right) \left(1 - \sqrt{\frac{q}{t}} Q
    q^{k} t^{j - i + 1 }\right)} =  G^{(q,t)}_{YW}\left(\sqrt{\frac{q}{t}} Q\right).\\
  M_Y^{(q,t)}(t^{-\rho}) M_{Y^{\mathrm{T}}}^{(t,q)}(q^{-\rho}) =
  \frac{(-1)^{|Y|}t^{\frac{|Y|}{2}} q^{-\frac{|Y|}{2}}}{G_{YY}^{(q,t)}
    (1)} \label{eq:3}
\end{gather}
Using these identities, the standard $q$-Selberg measure evaluated at
discrete points $x_{R,i} = q^{R_i+1}t^{N-i}$ can be expressed in
several convenient ways
\begin{multline}
\label{eq:17}
  \frac{\mu_{A_1}(x_R|u,v,N,q,t)}{\mu_{A_1}(x_{\varnothing}|u,v,N,q,t)} =
  (-1)^{|R|} q^{(u+v+1)|R|} M_R^{(q,t)} \left( \frac{1 -
      t^{nN}}{1 - t^n} \right) M_{R^{\mathrm{T}}}^{(t,q)}
  \left( \frac{1 - t^{-n (N-1)} q^{- n (v+1)}}{1 - q^n} \right) =\\
  = (-1)^{|R|} q^{(u+v+1/2)|R|} t^{-|R|/2} M_R^{(q,t)} \left(p_n(
    t^{-\rho} ) - p_n (t^N t^{-\rho}) \right)
  M_{R^{\mathrm{T}}}^{(t,q)} \left( p_n (q^{-\rho}) - p_n (t^{1-N}
    q^{-v-1} q^{-\rho} ) \right) =\\
  = \left( t q^{u-1} \right)^{|R|} \frac{G^{(q,t)}_{R\varnothing}(q^{v+1} t^{N-1}) G^{(q,t)}_{\varnothing
      R}(q t^{-N-1})}{G^{(q,t)}_{RR}(1)}=\\
  = (-1)^{|R|} q^{u|R|} t^{|R|} M_R^{(q,t)} \left( \frac{1 -
      q^{n(v+1)} t^{n(N-1)} }{1 - t^n} \right) M_{R^{\mathrm{T}}}^{(t,q)}
  \left( \frac{1 - t^{-n N}}{1 - q^n} \right)=\\
  = (-1)^{|R|} q^{(u-1/2)|R|} t^{-|R|/2} M_R^{(q,t)} \left(p_n(
    t^{-\rho} ) - p_n ( q^{v+1} t^{N-1} t^{-\rho}) \right)
  M_{R^{\mathrm{T}}}^{(t,q)} \left( p_n (q^{-\rho}) - p_n (t^{-N}
    q^{-\rho} ) \right)=\\
    = q^{(u+v)|R|} \frac{G^{(q,t)}_{R\varnothing}(t^N) G^{(q,t)}_{\varnothing
      R}(q^{-v} t^{-N})}{G^{(q,t)}_{RR}(1)}.
\end{multline}


\begin{thebibliography}{12}

  \bibitem{AGT} L.~Alday, D.~Gaiotto and Y.~Tachikawa,
  Lett.\ Math.\ Phys.\ {\bf 91} (2010) 167--197, arXiv:0906.3219\\
  N.~Wyllard,
  JHEP {\bf 0911} (2009) 002, arXiv:0907.2189\\
  A.~Mironov and A.~Morozov, Nucl.\ Phys.\ {\bf B825} (2009) 1--37,
  arXiv:0908.2569

\bibitem{5dAGT}
  H.~Awata and Y.~Yamada,
  JHEP {\bf 1001} (2010) 125,
arXiv:0910.4431;
  Prog.\ Theor.\ Phys.\  {\bf 124} (2010) 227,
arXiv:1004.5122\\
S. Yanagida, arXiv:1005.0216\\
F. Nieri, S. Pasquetti, F. Passerini and A. Torrielli, arXiv:1312.1294\\
H. Itoyama, T.Oota and R. Yoshioka, arXiv:1408.4216, arXiv:1602.01209\\
  A. Nedelin and M. Zabzine, arXiv:1511.03471\\
  R. Yoshioka, arXiv:1512.01084\\
    Y.~Ohkubo, H.~Awata and H.~Fujino,
  arXiv:1512.08016


\bibitem{6dAGT}
  A.~Iqbal, C.~Kozcaz and S.~T.~Yau,
  arXiv:1511.00458\\
F.~Nieri,
  arXiv:1511.00574

\bibitem{Gaiotto} D.Gaiotto, arXiv:0908.0307

\bibitem{Aharony:1997bh}
S.~H.~Katz, A.~Klemm and C.~Vafa,
  Nucl.\ Phys.\ B {\bf 497} (1997) 173, hep-th/9609239\\
  S.~Katz, P.~Mayr and C.~Vafa,
  Adv.\ Theor.\ Math.\ Phys.\  {\bf 1} (1998) 53, hep-th/9706110\\
  B. Kol, 
  JHEP {\bf 9911} (1999) 026, hep-th/9705031\\
  O.~Aharony, A.~Hanany and B.~Kol,
  JHEP {\bf 9801} (1998) 002, hep-th/9710116\\
  B. Kol and J. Rahmfeld, 
  JHEP {\bf 9808}, 006 (1998),
  hep-th/9801067

\bibitem{GGM2} A. Gorsky, S. Gukov and A. Mironov,
Nucl. Phys. {\bf B518} (1998) 689, arXiv:hep-th/9710239

\bibitem{ref} A. Iqbal, hep-th/0207114\\
M. Aganagic, A. Klemm, M. Marino and C. Vafa, Commun. Math. Phys. {\bf 254} (2005) 425 hep-th/0305132\\
M. Taki, JHEP {\bf 0803} (2008) 048, arXiv:0710.1776\\
 H. Awata and H. Kanno, JHEP {\bf 0505} (2005) 039, hep-th/0502061;
Int. J. Mod. Phys. {\bf A24} (2009) 2253,
arXiv:0805.0191

\bibitem{IKV} A. Iqbal, C. Kozcaz and C. Vafa, JHEP {\bf 0910} (2009) 069, hep-th/0701156

\bibitem{spedu}
E. Mukhin, V. Tarasov and A. Varchenko, 
math/0510364;
Adv. Math. {\bf 218} (2008) 216-265, math/0605172\\
A. Mironov, A. Morozov, Y. Zenkevich and A. Zotov, JETP Lett. {\bf 97} (2013) 45, arXiv:1204.0913\\
A. Mironov, A. Morozov, B. Runov, Y. Zenkevich and A. Zotov, Lett. Math. Phys. {\bf 103} (2013) 299,
arXiv:1206.6349; JHEP {\bf 1312} (2013) 034, arXiv:1307.1502

\bibitem{MMZZ} A. Mironov, A. Morozov, Y. Zenkevich and A. Zotov, to appear

\bibitem{Bao} L. Bao, E. Pomoni, M. Taki and F. Yagi, JHEP {\bf 1204} (2012) 105, arXiv:1112.5228

\bibitem{enhanced} M. Taki, arXiv:1310.7509\\
O. Bergman, D. Rodrigues-Gomez and G. Zafrir, JHEP {\bf 03} (2014) 112, arXiv:1311.4199\\
G. Zafrir, JHEP {\bf 12} (2014) 116, arXiv:1408.4040\\ 
O. Berman and G. Zafrif, JHEP {\bf 04} (2015) 141, 	arXiv:1410.2806\\
V. Mitev, E. Pomoni, M. Taki and F. Yagi, JHEP {\bf 04} (2015) 052, arXiv:1411.2450\\
S.-S. Kim, M. Taki and F. Yagi, Prog. Theor. Exp. Phys. (2015) 083B02, arXiv:1504.03672

\bibitem{Aganagic:2013tta} 
M.~Aganagic, N.~Haouzi, C.~Kozcaz and
  S.~Shakirov, 
  arXiv:1309.1687 \\
M.~Aganagic, N.~Haouzi and S.~Shakirov, 
  arXiv:1403.3657

\bibitem{Sulk}
P. Sulkowski and A. Klemm, Nucl. Phys. {\bf B819} (2009) 400-430, arXiv:0810.4944\\
P. Sulkowski, JHEP {\bf 04} (2010) 063, arXiv:0912.5476

\bibitem{FDagt} Vl. Dotsenko and V. Fateev, Nucl. Phys. {\bf B240}
(1984) 312-348\\
A. Marshakov, A. Mironov and A. Morozov,
Phys. Lett. {\bf B265} (1991) 99\\
S. Kharchev, A. Marshakov, A. Mironov, A. Morozov and S. Pakuliak,
Nucl. Phys. {\bf B404} (1993) 717-750,  hep-th/9208044\\
H. Awata, Y. Matsuo, S. Odake and J. Shiraishi, Soryushiron Kenkyu {\bf 91} (1995) A69-A75, hep-th/9503028\\
H. Itoyama, K. Maruyoshi and T. Oota,
Prog.Theor.Phys. {\bf 123} (2010) 957-987, arXiv:0911.4244\\
T. Eguchi and K. Maruyoshi,
arXiv:0911.4797;
arXiv:1006.0828\\
R. Schiappa and N. Wyllard,
arXiv:0911.5337\\
A. Mironov, A. Morozov and S. Shakirov, JHEP {\bf 1002} (2010) 030, arXiv:0911.5721; Int. J. Mod. Phys.
{\bf A25} (2010) 3173, arXiv:1001.0563; J. Phys. {\bf A44} (2011) 085401, arXiv:1010.1734; JHEP {\bf 1103} (2011)
102, arXiv:1011.3481; Int. J. Mod. Phys. {\bf A27} (2012) 1230001, arXiv:1011.5629\\
H. Itoyama and T. Oota, Nucl. Phys. {\bf B838} (2010) 298-330, arXiv:1003.2929\\
A. Mironov, A. Morozov, and And. Morozov, Nucl. Phys. {\bf B843} (2011) 534, arXiv:1003.5752

\bibitem{DV} R. Dijkgraaf and C. Vafa, arXiv:0909.2453

\bibitem{MMSS5} A. Mironov, A. Morozov, S. Shakirov and A. Smirnov, Nucl. Phys. {\bf B855} (2012) 128, arXiv:1105.0948

\bibitem{genMD}
  Y.~Zenkevich, 
  JHEP {\bf 1505} (2015) 131, arXiv:1412.8592

\bibitem{Morozov:2015xya} A.~Morozov and Y.~Zenkevich,
JHEP {\bf 1602} (2016) 098,  arXiv:1510.01896

\bibitem{MMZ} A.~Mironov, A.~Morozov, Y.~Zenkevich, 
    arXiv:1512.06701

\bibitem{GKMMM} A.Gorsky, I.Krichever, A.Marshakov, A.Mironov, A.Morozov,
Phys.Lett. {\bf B355} (1995) 466, hep-th/9505035

\bibitem{SWint} E. Martinec,
Phys. Lett. {\bf B367} (1996) 91-96, hep-th/9510204\\
E. Martinec and N. Warner,
Nucl. Phys. {\bf 459} (1996) 97, hep-th/9511052\\
I.M. Krichever and D.H. Phong, J. Diff. Geom. {\bf 45} (1997) 349-389, hep-th/9604199

\bibitem{SWCal} R. Donagi and E. Witten, Nucl. Phys. {\bf B460} (1996) 299-334, hep-th/9510101\\
H. Itoyama and A. Morozov,
Nucl. Phys. {\bf B477} (1996) 855-877, hep-th/9511126;
Nucl. Phys. {\bf B491} (1997) 529-573, hep-th/9512161

\bibitem{XXX} A. Gorsky, A. Marshakov, A. Mironov, A. Morozov,
Phys. Lett. {\bf B380} (1996) 75-80, arXiv:hep-th/9603140

\bibitem{XYZ} A. Gorsky, A. Marshakov, A. Mironov and A. Morozov, hep-th/9604078

\bibitem{GGM1} A. Gorsky, S. Gukov and A. Mironov,
Nucl. Phys. {\bf B517} (1998) 409-461, hep-th/9707120

\bibitem{MirMar} A. Marshakov and A. Mironov,
Nucl. Phys. {\bf B518} (1998) 59-91, hep-th/9711156

\bibitem{Nek5} N. Nekrasov,
Nucl. Phys. {\bf B531} (1998) 323-344, hep-th/9609219

\bibitem{Braden} H. W. Braden, A. Marshakov, A. Mironov and A. Morozov, Phys. Lett. {\bf B448} (1999) 195, hep-th/9812078;
Nucl. Phys. {\bf B558} (1999) 371, hep-th/9902205

\bibitem{GMABCD} A. Gorsky and A. Mironov, Nucl. Phys. {\bf B550} (1999) 513, hep-th/9902030

\bibitem{Dell} H. W. Braden, A. Marshakov, A. Mironov, A. Morozov,
  Nucl. Phys. {\bf B573} (2000) 553--572, hep-th/9906240\\
  A. Mironov and A. Morozov, Phys. Lett. {\bf B475} (2000) 71-76, hep-th/9912088; hep-th/0001168\\
  G.~Aminov, A.~Mironov, A.~Morozov and A.~Zotov,
  Phys. Lett. {\bf B726} (2013) 802, arXiv:1307.1465\\
  G.~Aminov, H.~W.~Braden, A.~Mironov, A.~Morozov and A.~Zotov,
  JHEP {\bf 1501} (2015) 033 arXiv:1410.0698

\bibitem{NSM} N. Nekrasov and S. Shatashvili, arXiv:0908.4052\\
A. Mironov and A. Morozov, 
JHEP {\bf 04} (2010) 040, arXiv:0910.5670;
J. Phys. {\bf A43} (2010) 195401, arXiv:0911.2396

\bibitem{GM} A. Gorsky and A. Mironov,
hep-th/0011197

\bibitem{Iqbal:2003ds} A.~Iqbal, N.~Nekrasov, A.~Okounkov and C.~Vafa,
  JHEP {\bf 0804} (2008) 011,
hep-th/0312022

\bibitem{Nekr-QUARKS} N.~Nekrasov, \textit{Instanton Partition
    Functions and M-Theory,} Proceedings of 15$^{\text{th}}$
  International Seminar on High Energy Physics (Quarks 2008)

\bibitem{Nekrasov:2014nea} N.~Nekrasov and A.~Okounkov,
  arXiv:1404.2323

\bibitem{LMNS}
G. Moore, N. Nekrasov and S. Shatashvili, Nucl. Phys. {\bf B534} (1998) 549-611, hep-th/9711108;
hep-th/9801061\\
A. Losev, N. Nekrasov and S. Shatashvili, Commun. Math. Phys. {\bf 209} (2000) 97-121, hep-th/9712241;
ibid. 77-95, hep-th/9803265

\bibitem{Witten:1997sc}
  E.~Witten,
  Nucl.\ Phys.\ B {\bf 500} (1997) 3, 
hep-th/9703166

\bibitem{Witmore} 
D.-E.Diaconescu,
Nucl. Phys. {\bf B503} (1997) 220-238, hep-th/9608163\\
A.Hanany and E.Witten,
Nucl. Phys. {\bf B492} (1997) 152, hep-th/9611230\\
J. de Boer, K. Hori, H. Ooguri, Y. Oz and Z. Yin, 
Nucl. Phys. {\bf B493} (1997) 148-176, hep-th/9612131\\
J. de Boer, K. Hori, Y. Oz and Z. Yin, 
Nucl. Phys. {\bf B502} (1997) 107-124, hep-th/9702154\\
S. Elitzur, A. Giveon and D. Kutasov, 
Phys. Lett. {\bf B400} (1997) 269-274, hep-th/9702014\\
A. Marshakov, A. Morozov and M. Martellini, Phys. Lett. {\bf B418} (1998) 294-302, hep-th/9706050

\bibitem{Leung:1997tw}
  N.~C.~Leung and C.~Vafa,
  Adv.\ Theor.\ Math.\ Phys.\  {\bf 2} (1998) 91, hep-th/9711013

\bibitem{CB} A. Belavin, A. Polyakov and A. Zamolodchikov, Nucl. Phys. {\bf B241} (1984) 333-380\\
A. Zamolodchiková Al. Zamolodchikov, Conformal field theory and critical phenomena in 2d systems, 2009\\
L. Alvarez-Gaume, Helvetica Physica Acta {\bf 64} (1991) 361\\
P. Di Francesco, P. Mathieu and D. Senechal, Conformal Field Theory, Springer, 1996\\
A. Mironov, S. Mironov, A. Morozov and An. Morozov, Theor. Math. Phys. {\bf 165} (2010) 1662-1698, arXiv:0908.2064

\bibitem{SW} N. Seiberg and E. Witten,
Nucl. Phys. {\bf B426} (1994) 19-52, hep-th/9407087;
Nucl. Phys. {\bf B431} (1994) 484-550, hep-th/9408099

\bibitem{Nekrasov:2002qd}
N. Nekrasov, Adv. Theor. Math. Phys. {\bf 7} (2004) 831-864, hep-th/0206161\\
R. Flume and R. Pogossian, Int. J. Mod. Phys. {\bf A18} (2003) 2541\\
N. Nekrasov and A. Okounkov, hep-th/0306238

\bibitem{MMMselb} A.~Mironov, A.~Morozov, and And.~Morozov,
  Nucl.\ Phys.\ {\bf B843} (2011) 534, arXiv:1003.5752

\bibitem{MMSha} A. Mironov, A. Morozov and S. Shakirov, JHEP {\bf 1102} (2011) 067,
arXiv:1012.3137

\bibitem{prev1} 
A.~Morozov and A.~Smirnov,
   Lett.\ Math.\ Phys.\ {\bf 104} (2014) 585, arXiv:1307.2576 \\
   S.~Mironov, An.~Morozov and Y.~Zenkevich,
  JETP Lett.\  {\bf 99} (2014) 109, arXiv:1312.5732 \\
  Y.~Ohkubo, arXiv:1404.5401\\
A.~Smirnov,  arXiv:1404.5304;  arXiv:1302.0799\\
  B.~Feigin, M.~Jimbo, T.~Miwa and E.~Mukhin, arXiv:1502.07194 

\bibitem{prev2} M.C.N. Cheng, R. Dijkgraaf and C. Vafa, JHEP {\bf 09} (2011) 022, arXiv:1010.4573\\
M. Aganagic, M.C.N. Cheng, R. Dijkgraaf, D. Krefl and C. Vafa, JHEP {\bf 11} (2012) 019, 	arXiv:1105.0630  

\bibitem{Aganagic:2015cta}
  M.~Aganagic and N.~Haouzi,
  arXiv:1506.04183

\bibitem{Kimura:2015rgi}
  T.~Kimura and V.~Pestun,
  arXiv:1512.08533

   \bibitem{Haghighat:2013gba}
B.Haghighat, A.Iqbal, C.Kozcaz, G.Lockhart, C.Vafa,
  Comm.Math.Phys. {\bf 334} (2015) 779,
arXiv:1305.6322

  B.~Haghighat, C.~Kozcaz, G.~Lockhart and C.~Vafa,
  Phys.\ Rev.\ {\bf D89} (2014) 4, 046003,
arXiv:1310.1185

  
\bibitem{Nakamura:2014nha}
  S.~Nakamura, F.~Okazawa and Y.~Matsuo,
  Prog. Theor. Exp. Phys. {\bf 3} (2015) 033B01,
  arXiv:1411.4222\\
  S.~Nakamura,
  Prog. Theor. Exp. Phys. {\bf 2015} (2015) 7, 
  arXiv:1502.04188\\
  S. Kanno, Y. Matsuo and H. Zhang, arXiv:1306.1523

\bibitem{Szabo:2015wua}
  H.~Awata and H.~Kanno,
  JHEP {\bf 0907} (2009) 076,
arXiv:0905.0184\\
  M.~Cirafici, A.~Sinkovics and R.~J.~Szabo,
  Nucl.\ Phys.\ B {\bf 853} (2011) 508,
arXiv:1012.2725\\
R.~J.~Szabo,
arXiv:1507.00685

\bibitem{HV} K. Hasegawa, J. Phys. {\bf A26} (1993) 3211-3228; Comm. Math. Phys. {\bf 187} (1997) 289, q-alg/9512029\\
V. Vakulenko, math/9909079

\bibitem{Skl} E. Sklyanin, Func. Anal. \& Apps. {\bf 16} (1982) 27; {\bf 17} (1983) 34

\bibitem{Rui} S.N.M. Ruijsenaars and H. Schneider,
Ann. Phys. (NY) {\bf 170} (1986) 370\\
S.N.M. Ruijsenaars, Comm. Math. Phys. {\bf 110} (1987) 191-213

\bibitem{KZ} I. Krichever and A. Zabrodin, Uspekhi Mat. Nauk {\bf 50} (1995) 3-56, hep-th/9505039 

\bibitem{mCal} N. Nekrasov, Commun. Math. Phys. {\bf 180} (1996) 587-604, hep-th/9503157\\
A.M. Levin, M.A. Olshanetsky and A. Zotov, Comm. Math. Phys. {\bf 236} (2003) 93–133, nlin/0110045

\bibitem{AHZ} A. Antonov, K. Hasegawa and A. Zabrodin, Nucl. Phys. {\bf B503} (1997) 747-770, hep-th/9704074

\bibitem{OF} B.L. Feigin and A. V. Odesskii, 
Funct. Anal. Appl. {\bf 23} (1989) 45-54; {\bf 27} (1993) 31-38; math/9812059\\
H.W. Braden, V.A. Dolgushev, M.A. Olshanetsky and A.V. Zotov,
J. Phys. {\bf A36} (2003) 6979-7000, hep-th/0301121\\
A. Odesskii and V. Rubtsov, math/0110032; math/0404159

\bibitem{amitoappear} G. Aminov, A. Mironov and A. Morozov, to appear

\bibitem{Nek} E. Carlsson, N. Nekrasov and A. Okounkov, arXiv:1308.2465

\end{thebibliography}
\end{document}